\documentclass[fleqn,usenatbib]{mnras}

\usepackage{newtxtext,newtxmath}
\usepackage[T1]{fontenc}

\DeclareRobustCommand{\VAN}[3]{#2}
\let\VANthebibliography\thebibliography
\def\thebibliography{\DeclareRobustCommand{\VAN}[3]{##3}\VANthebibliography}

\usepackage{graphicx}	% Including figure files
\usepackage{amsmath}	% Advanced maths commands
\usepackage{gensymb}
\usepackage{lineno}

\title[Mantle Convection and Volcanism on K2-141\,b]{Mantle Convection and Nightside Volcanism on Lava World K2-141\,b}

\author[Tobias G. Meier et al.]{
Tobias G. Meier,$^{1,2}$\thanks{E-mail: tobias.meier@physics.ox.ac.uk}
Claire Marie Guimond,$^{1}$
Raymond T. Pierrehumbert,$^{1}$
Jayne Birkby, $^{2}$
\newauthor{
Richard D. Chatterjee,$^{1,6}$
Chloe E. Fisher,$^{2}$
Gregor J. Golabek,$^{3}$
Mark Hammond,$^{1}$
Thaddeus D. Komacek$^{1}$
}
\newauthor{
Tim Lichtenberg,$^{4}$
Alex McGinty,$^{1}$ 
Erik Meier Vald\'es,$^{2}$
Harrison Nicholls,$^{1}$
Luke T. Parker,$^{2}$
}
\newauthor{
Rob J. Spaargaren,$^{4}$
Paul J. Tackley$^{5}$
}
\\
$^{1}$ Atmospheric, Oceanic and Planetary Physics, Department of Physics, University of Oxford, Parks Road, Oxford OX1 3PU, UK\\
$^{2}$ Astrophysics, University of Oxford, Denys Wilkinson Building, Keble Road, Oxford, OX1 3RH, UK \\
$^{3}$ Bayerisches Geoinstitut, University of Bayreuth, Universitätsstrasse 30, 95440 Bayreuth, Germany \\
$^{4}$ Kapteyn Astronomical Institute, University of Groningen, PO Box 800, 9700 AV Groningen, The Netherlands \\
$^{5}$ Institute of Geophysics, Department of Earth and Planetary Sciences, ETH Zurich, Sonneggstrasse 5, Zurich 8092, Switzerland \\
$^{6}$ School of Physics and Astronomy, University of Leeds, LS2 9JT, UK
}
\date{Accepted XXX. Received YYY; in original form ZZZ}

\pubyear{\the\year{}}

\begin{document}
%\linenumbers
\label{firstpage}
\pagerange{\pageref{firstpage}--\pageref{lastpage}}
\maketitle

\begin{abstract}
Ultra-short period lava worlds offer a unique window into the coupled evolution of planetary interior and atmospheres under extreme irradiation. In this study, we investigate the mantle dynamics, nightside volcanism, and volatile outgassing on lava world K2-141\,b ($1.54\,R_\oplus$, $5.31\,M_\oplus$) using two-dimensional convection models with tracer-based volatile tracking. Our simulations explore a range of interior configurations, including models with and without plastic yielding, basal versus mixed heating, core cooling, and melt intrusion. 
In models without plastic yielding (i.e. with a strong lithosphere), we find that mantle upwellings form at the substellar and antistellar points, while downwellings form near the day-night terminators at the boundary between the magma ocean and cold, solid nightside. These downwellings facilitate the recycling of crustal material, representing a form of asymmetric, single-lid tectonics. The resulting magma ocean thickness varies from $200$ to $300$\,km depending on the model parameters, corresponding to about $2$-$3\%$ of the planet's radius.  
Continuous nightside volcanism produces a basaltic crust and gradually depletes the mantle of volatiles. We find that over a billion years, volcanic eruptions can outgas tens of bars of CO$_2$ and H$_2$O. We show that even relatively large volcanic eruptions on the nightside produce thermal emission signals of no more than 1\,ppm, remaining below the current detectability threshold in thermal phase curves. However, for most models, outgassing rates are increased near the day-night terminators and future studies should assess whether such localised outgassing could lead to atmospheric signatures in transmission spectroscopy. 
\end{abstract}


%%%%%%%%%%%%%%%%%%%%%%%%%%%%%%%%%%%%%%%%%%%%%%%%%%

%%%%%%%%%%%%%%%%% BODY OF PAPER %%%%%%%%%%%%%%%%%%

\section{Introduction}
Many rocky exoplanets orbit their host star with ultra-short periods of less than one day. Due to these short orbital periods, these planets are likely to be tidally locked \citep{Peale1977, Lyu2024}, resulting in permanent dayside and nightside hemispheres. For some of these exoplanets, the intense instellation they receive from the host star is sufficient to melt, and potentially evaporate, their surfaces, forming magma oceans. These so-called lava worlds \citep{Leger2011, Chao2021} have recently gained significant attention in the search for atmospheres around rocky exoplanets \citep{lichtenberg2025}. A possible CO/CO$_2$ atmosphere has recently been reported for 55 Cancri e \citep{Hu2024}, a lava world with a radius of $1.88\pm0.03\,R_\oplus$ and mass of $8\pm0.3\,M_\oplus$ \citep{McArthur2004, Bourrier2018}. However, follow-up observations will be necessary to confirm such a volatile atmosphere as observations have also shown strong variability in the observed eclipse depth, which has prevented a definitive characterisation \citep{Patel2024}. 

If volatiles are in chemical and vapour-pressure equilibrium between the magma ocean and an overlying atmosphere, then characterising the chemical species in the atmosphere can offer direct insights into the interior composition and outgassing history of rocky exoplanets \citep[e.g.][]{Lichtenberg2021, nicholls_convective_2025}. However, the volatile inventory and outgassing efficiency of the magma ocean is itself influenced by interior processes, such as the convective vigour of the magma ocean \citep{Salvador2023, Walbecq2025}, redox reactions that could lead to a compositionally stratified mantle that is stable against convection \citep{ModirroustaGalian2025}, or the delivery of volatiles from the mantle itself. To correctly interpret current and future atmospheric observations of tidally locked rocky exoplanets, it is therefore essential to understand how tidal locking and permanent dayside magma oceans influence mantle convection regimes and the associated transport and outgassing of volatiles. 
Recent modelling efforts have highlighted the complex interplay between interior dynamics, magma ocean evolution, and volatile outgassing on lava worlds. \citet{Boukare2025} performed 2D and 3D simulations of magma ocean circulation and showed that strong surface temperature gradients can drive asymmetric global circulation patterns with nightside downwellings and a return flow on the dayside. They propose two end-member regimes for lava planets: Either a hot, fully molten interior with a shallow nightside crust where the atmosphere is in equilibrium with the planet's bulk-silicate composition, or a cold solid-state regime where the atmosphere is in equilibrium with a thin, fractionated dayside magma ocean. For planets with negligible internal heat sources, \cite{Lai2024} showed that the magma oceans are much shallower than previously estimated from adiabatic models, which could potentially reduce the efficiency of volatile exchange between the interior and atmosphere. It has been further demonstrated that shallow magma oceans tend to be more reduced--particularly on smaller planets--while deeper magma oceans on larger planets outgas more oxidised species \citep{Armstrong2019, Deng2020, Maurice2024}. 

While many recent studies have focused on the equilibrium between the magma ocean and overlying atmosphere \citep[e.g.][]{Hirschmann2012, Sossi2020, Gaillard2021, Bower2022, Nicholls2024}, the role of solid state mantle dynamics remains poorly understood. In tidally locked planets, intense stellar irradiation leads to strong day-night temperature contrasts, which can create asymmetries in tectonic regimes \citep{Gelman2011, VanSummeren2011, Meier2021, Meier2024}. A potential dayside magma ocean might however not extend across the entire surface, particularly if the planet lacks an atmosphere capable of efficiently redistributing heat from the intensely irradiated dayside to the nightside \citep{Kite2016}. In the absence of such redistribution, the nightside could reach extremely low temperatures (below $100$\,K). \cite{Meier2023} investigated how the large day-night surface temperature contrast on 55 Cancri e affects mantle dynamics and found that mantle convection is strongly asymmetric, with super-plumes rising preferentially on the dayside from the core-mantle-boundary (CMB) towards the dayside magma ocean if the nightside remains solid. Conversely, if the planet has a substantial atmosphere, the nightside temperature can be significantly higher \citep{Hammond2017}. The interaction between the convecting solid mantle and the lower boundary of the magma ocean could influence the extent of the magma ocean and the supply of volatiles \citep[e.g.][]{Meier2023, Salvador2023, Walbecq2025}. 
While the surface temperature contrast shapes the thermal boundary condition at the top of the mantle, internal heat sources are equally important in governing the planet's thermal and volcanic activity. In this study, we define volcanism as the generation of partial melt within the mantle that is transported to the surface. 
The mantle loses heat primarily through mantle convection, driven by several internal heat sources, such as radiogenic heating, residual heat from the planet's formation \citep[e.g.][]{ElkinsTanton2012,Solomatov2015}, and tidal dissipation \citep[e.g.][]{Hay2019, Bolmont2020, Nicholls2025}. 

The planet's cooling history and tectonic regime, and mantle composition control how efficiently melt is generated and transported within the mantle, thereby influencing the intensity and distribution of volcanic activity \citep{Noack2017, Dorn2018, Dehant2019, Spaargaren2020, Guimond2021, Lourenco2023}.
To explore the sensitivity of volcanic activity to interior heating, we consider two scenarios: one in which the mantle is purely basally heated---i.e. by the heat flux originating from the planet's core---and another that includes a combination of basal and internal heating, analogous to Earth's present-day mantle. The purely basally heated case may be relevant for planets that formed with a depleted inventory of heat-producing elements, such as young or volatile-poor planets \citep{Frank2014}.

Volcanism is a key process linking mantle convection to atmospheric processes on rocky planets. Recent observations of the sub-Earth L98-59\,b have sparked interest by revealing possible SO$_2$ features \citep{BelloArufe2025}, which has previously been proposed as a promising spectroscopic tracer for volcanic activity \citep{Kaltenegger2010, Seligman2024}. 
However, recent coupled interior-atmosphere models show that the volcanic activity required to maintain such an atmosphere on L98-59\,b---when balanced against escape---would require extreme levels of outgassing over Gyr timescales \citep{Nicholls2025}. Within our solar system, Io demonstrates that intense tidal heating can drive continuous global volcanism \citep{Spencer2000,Davies2003}, and recent Magellan SAR analyses of Venus have shown evidence of lava flows linked to volcanic activity during the spacecraft's mission from $1990$ to $1992$. These examples illustrate that diverse geodynamic settings can give rise to active volcanism. 

It has recently been shown that mid-infrared spectra obtained with JWST could be used to distinguish between basaltic surfaces with different mineralogical and compositional characteristics on airless rocky exoplanets \citep{First2025}. Moreover, \citet{Hammond2025} have shown that surface and atmospheric properties can be highly degenerate in thermal emission spectra, emphasising the need for for coupled interior-surface-atmosphere models to interpret thermal emission spectra. Consequently, interpreting atmospheric and surface observations of rocky exoplanets requires a detailed understanding of how geodynamic processes govern both volcanic outgassing and the evolution of surface composition.

To investigate how extreme surface temperature contrasts on tidally locked lava worlds influence interior dynamics and volatile outgassing, we present a set of two-dimensional mantle convection simulations of lava world K2-141\,b. This well-characterised ultra-short period exoplanet has a measured radius of $1.54^{+0.10}_{-0.09}\,R_{\earth}$ and mass of $5.31 \pm 0.46\,M_{\earth}$ \citep{Barragan2018}, yielding a bulk density that is consistent with that of a rocky planet without a significant volatile envelope. The Spitzer phase curve revealed a dayside brightness temperature of $T_\text{day}=2049^{+362}_{-359}$\,K in the $4.5\mu\,$m and a nightside temperature that is consistent with zero kelvin. If the planet has a sufficiently thin and transparent atmosphere, these brightness temperatures provide estimates for the surface temperature of the planet. A thermal phase curve of K2-141\,b has recently been obtained with JWST, which will provide new insights into the planet's atmospheric and surface properties \citep{Dang2021}. 

To capture the interplay between interior dynamics and surface conditions on such a planet, we perform mantle convection simulations using \textsc{StagYY} \citep{Tackley2008}, incorporating a thermal upper boundary condition informed by the Spitzer observations. We explore how different rheologies (with and without plastic yielding), internal heating (basal vs mixed), and melt emplacement style (intrusive vs extrusive) affect not only the thermal evolution but also the distribution of crust formation and volcanic outgassing. We use Lagrangian tracer particles to explicitly track melt generation, crustal composition, and the transport of volatile species (CO$_2$, H$_2$O) throughout the mantle and to the surface. This enables us to quantify where and when volcanic outgassing occurs, and how it may shape the chemical and thermal evolution of the planet. We aim to assess whether volcanic activity and volatile release can occur on the cold, solid nightside of K2-141\,b. We evaluate both hydrodynamic (thermal) and non-thermal escape channels, highlighting the role a planetary magnetic field could play in suppressing escape from a super-Earth under such intense irradiation.  By examining the links between mantle convection patterns, surface composition, and outgassing behaviour, our models offer new insights into the thermochemical evolution of tidally locked lava worlds and the prospects for detecting volcanic or atmospheric signatures on these planets. 

The outline of this paper is as follows: in Section~\ref{sec:methods}, we describe the numerical setup used to simulate mantle convection and volcanic outgassing on K2-141\,b. Section~\ref{sec:results} presents the results of our simulations focussing on the thermal evolution, convection regimes, heat flux, and evolution of the surface composition. In Section~\ref{sec:discussion}, we discuss the implications of our findings for the long-term evolution of lava worlds, including magma ocean thickness, thermal emission from the nightside, and retention of volatiles from volcanic outgassing. 

\section{Methods} \label{sec:methods}
\subsection{Model setup}
K2-141\,b has a measured radius of $1.54^{+0.10}_{-0.09}\,R_{\earth}$ and mass of $5.31 \pm 0.46\,M_{\earth}$, which is consistent with a planet that has a rocky composition lacking a thick atmosphere \citep{Barragan2018, Malavolta2018}. For simplicity, we therefore assume here that the core-to-planet radius ratio is the same as for Earth ($R_\mathrm{core}/R_{\mathrm{p}} \approx 0.55$). The planet orbits its host star on an ultra-short $6.7$\,hour orbit. Observations by Spitzer indicate a brightness dayside temperature of $2049^{+362}_{-359}$\,K with no measured thermal emission from the nightside, suggesting the presence of a dayside magma ocean and a cooler, non-molten nightside surface \citep{Zieba2022}. 
Age estimates for the host star K2-141 vary widely from $700\pm360$\,Myr to $6.3^{+6.6}_{-4.7}$\,Gyr \citep{Barragan2018, Malavolta2018}. In this study, we simulate ages up to $6.5$\,Gyr as a representative time scale within this range. This duration is sufficient for most models to reach a statistical steady state, in terms of mantle cooling, magma ocean depth, and the spatial distribution of plumes and downwellings. 
We used the thermochemical mantle convection code \textsc{StagYY} \citep{Tackley2008} in a two-dimensional spherical annulus geometry \citep{Hernlund2008} to solve for the mantle convection of K2-141\,b. We employed the truncated anelastic liquid approximation (TALA) with a reference density profile that varies with depth. The reference density profile was calculated using a 3\textsuperscript{rd}-order Birch-Murnaghan equation of state, with parameters for different mantle layers (upper mantle, transition zone, perovskite, and post-perovskite) chosen to reproduce the PREM density profile \citep{Dziewonski1981},  as in \cite{Meier2023}. For the thermal expansivity $\alpha$ and thermal conductivity $k$ we used the same parameters as in \cite{Tackley2013}. The models have a resolution of $64$\,cells in the vertical direction and $1024$\,cells in the angular direction. Similar to the study by \cite{Meier2023}, we assumed that diffusion creep is the dominant deformation mechanism and employed an Arrhenius-like viscosity law
\begin{equation}
\eta(P,T) = \eta_0\exp{\left(\frac{E_{\rm a}+PV_{\rm a}(P)}{RT}-\frac{E_{\rm a}}{RT_{\rm 0}}\right)}\,, 
\end{equation}
where $P$ is pressure, $T$ is temperature, $T_{\rm 0}$ is the temperature at $P=0$, $E_{\rm a}$ is the activation energy, $V_{\rm a}(P)$ is the activation volume and $R=8.31445$\,J\,K$^{-1}$\,$\mathrm{mol^{-1}}$ is the universal gas constant, and $\eta_{\rm 0}$ is a reference viscosity. The rheological parameters used are the same as those listed in Table 2 (lower bound post-perovskite) of \cite{Tackley2013} and as described in \cite{Meier2023}. 
To ensure numerical stability, we used minimum and maximum viscosity cut-offs of $10^{16}$\,Pa\,s, and $10^{28}$\,Pa\,s, respectively.
For some models, we also used a plastic yielding criteria to model the strength of the lithosphere where the yield stress $\sigma_{\rm y}$ is given by 
\begin{equation}
\sigma_{\mathrm{y}} = \mathrm{min}(c + c_{\rm f}P,\sigma_{\mathrm{duct}}+\sigma^{\prime}_{\mathrm{duct}}P)\,,
\end{equation}
where $c$ is the cohesive strength, $c_{\rm f}$ is the friction coefficient, $\sigma_\mathrm{duct}$ the ductile yield stress at $P=0$, and $\sigma^{\prime}_{\mathrm{duct}}$ is the ductile yield stress gradient.
If the stress exceeds the yield stress $\sigma_{\rm y}$, the viscosity gets reduced to an effective viscosity given by
\begin{equation} \label{eq:eta_eff}
\eta_{\text{eff}} = \frac{\sigma_{\text{y}}}{2 \dot{\varepsilon}_{\text{II}}} \quad \mathrm{if} \hspace{0.5em} 2 \eta \dot{\varepsilon}_{\text{II}} > \sigma_{\rm y}\,,
\end{equation}
where $\dot{\varepsilon}_{\text{II}}$ is the second invariant of the strain rate tensor. The corresponding values used for these parameters are shown in Table~\ref{tab:parameters}.
%%%%%%%%%%%%%%%%
%%% TABLE 1 %%%
\begin{table}
	\centering
	\caption{Model parameters}
	\label{tab:parameters}
	\begin{tabular}{lccr} % four columns, alignment for each
		\hline
		Property & Value & Symbol & Units\\
		\hline
		Mantle depth & 4450 & $d_{\text{mantle}}$ & km\\
        Core radius &  5359  & $R_{\text{core}}$  & km  \\
		Dayside temperature  & 3000 & $T_{\text{day}}$ & K\\
		Nightside temperature & 50 &  $T_{\text{night}}$ & K\\
        Internal heating rate &$5.2 \cdot 10^{-12}$ & $H$ & W\,kg$^{-1}$ \\ 
        CMB temperature &  6500  & $T_{\text{CMB}}$ & K \\
        Cohesive strength & $1$ & $c$ &  MPa\\
        Friction coefficient & $0.1$ & $c_{\rm f}$ & \\
        Ductile yield stress & $100$ & $\sigma_\mathrm{duct}$ & MPa \\
        Ductile yield stress gradient & $0.01$ & $\sigma^{\prime}_{\mathrm{duct}}$ & \\
		\hline
	\end{tabular}
\end{table}
%%%%%%%%%%%%%%%%
%%% END OF TABLE 1 %%%
For all models (see Table~\ref{tab:model_overview}), we set the initial CMB temperature to an initial value of $T_\mathrm{CMB}=6500$\,K. For some models (MixCC, MPlastic), we also include the effect of core cooling on the CMB temperature. For these cases, the CMB temperature decreases with time according to the averaged heat flux out of the core, assuming a heat capacity for the core of $C_\mathrm{p,core}=750$\,J\,kg$^{-1}$\,K$^{-1}$ \citep{Gubbins2003, Labrosse2014}. The models with mixed heating (MixCC, MixIntr) additionally include a uniform internal heating rate of $H=5.2 \cdot 10^{-12}$\,W\,kg$^{-1}$, corresponding to present-day Earth-like internal heating. 
The mechanical boundary conditions at both the surface and the CMB are set to be free-slip.
The surface temperature boundary condition is tidally-locked and radiative. The surface temperature is calculated locally from a balance between the incoming and outgoing radiative fluxes, given by
\begin{equation}
\sigma T_s^4 = F_{\mathrm{top}} + \sigma T_{\mathrm{rad}}^{4}\,,
\end{equation}
where $T_s$ is the surface temperature, $F_{\mathrm{top}}$ is the surface heat flux from the interior, $\sigma = 5.670 \cdot 10^{-8}$\,Wm$^{-2}$K$^{-4}$ is the Stefan-Boltzmann constant, and $T_\mathrm{rad}$ is the longitudinally varying equilibrium temperature imposed by synchronous rotation. 
\subsection{Melting and outgassing} \label{section:methods_melting}
\begin{figure}
    \centering
    \includegraphics[width=0.45\textwidth]{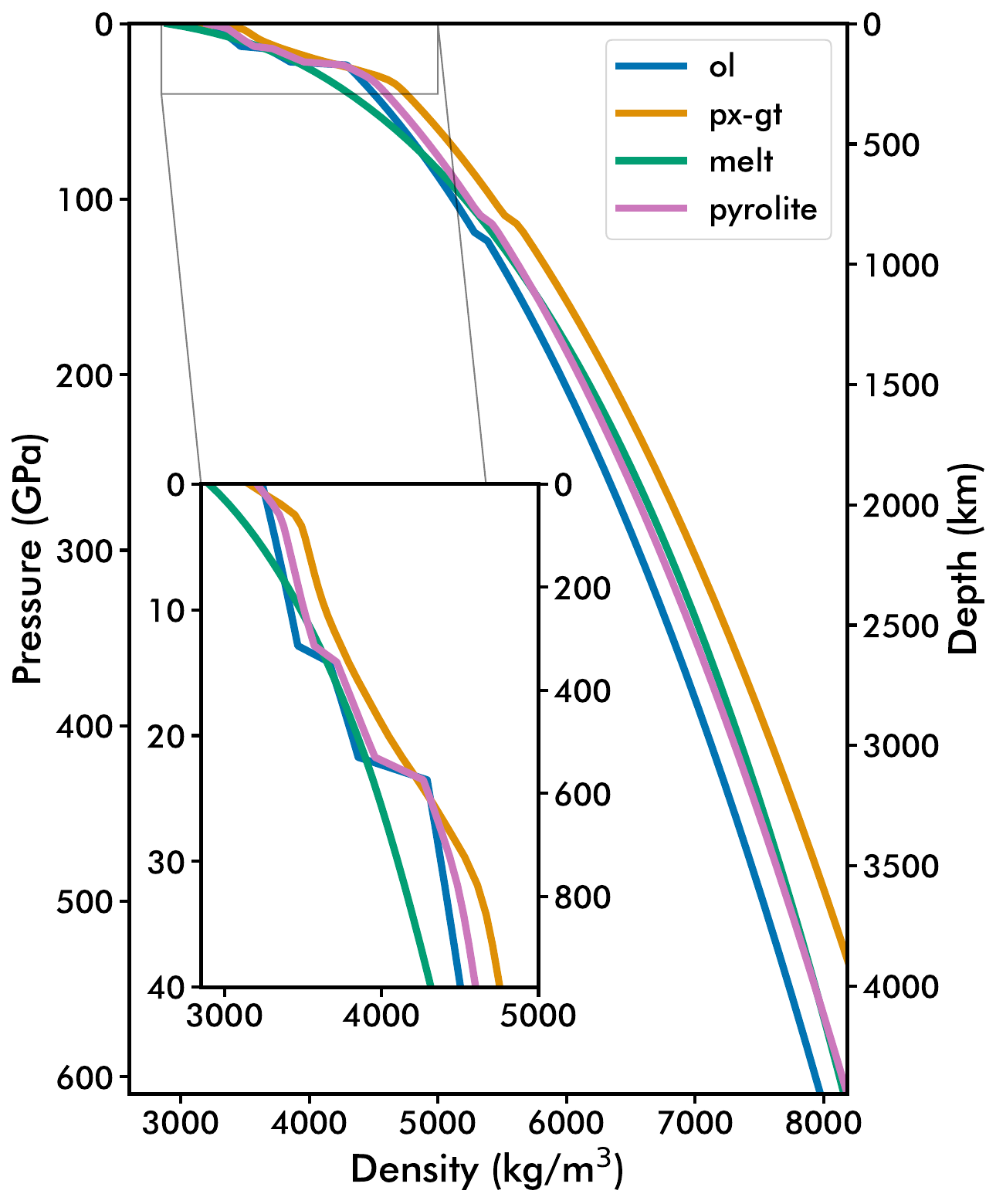}
    \caption{Reference density profile for lava world K2-141\,b including olivine (ol), pyroxene-garnet (px-gt), and melt (same for molten olivine and pyroxene-garnet). The purple line shows the combined reference density for pyrolite.}
    \label{fig:ref_density}
\end{figure}

We use Lagrangian tracers to track melt, composition, and volatiles in the mantle \citep[][]{Tackley2002, Tackley2003, Xie2004}. At the start of the model run, each cell was initialised with an average of $15$ tracers. Each tracer carries information on its position, type (i.e. molten/solid; see below), mass, and the concentrations of carbon and water.
The mantle composition of rocky planets is heterogeneous, consisting of various materials that melt at different temperatures. In this study, we adopt the pyrolite model \citep{Ringwood1962}, which is a theoretical model for Earth's upper mantle composition, which represents the bulk mantle as a mixture of approximately $20\%$ basalt and $80\%$ harzburgite. This corresponds to the primitive mantle material. This choice is motivated by the fact that the Solar System is typical in terms of rock-forming element abundances; very exotic mantle compositions are not expected to be common \citep{Spaargaren2023, Guimond2024}.
Through partial melting, basalt is extracted and forms basaltic crust (similar to oceanic crust) whereas harzburgite represents the depleted, solid mantle. We track this composition of the mantle by attributing each tracer a type, distinguishing between solid and molten basalt and harzburgite. All tracers are initialised as pyrolite ($0.2$ basalt, $0.8$ harzburgite fraction). 
After each time step, a grid-based composition is calculated from the tracers' average composition within each cell. We model the composition of a cell using a four-component system, comprising the end-member phases olivine (including three solid-solid phase transitions), a mixture of pyroxene/garnet (including four solid-solid phase transitions), molten olivine, and molten pyroxene/garnet. We assume that harzburgite is composed of $60\%$ olivine and $40\%$ pyroxene-garnet, while basalt is modelled using the same reference density profile as pyroxene/garnet. The reference density profiles for the end-member phases are shown in Figure~\ref{fig:ref_density}. We assume the same density profile for molten olivine and molten pyroxene-garnet. 
We use the same solidus and liquidus as in \cite{Meier2023}, where the pressure-dependent melting curves are shown in their Figure~A.1. The solidus is given by \cite{Herzberg2000} for the upper mantle, \cite{Zerr1998} for the lower mantle, and \cite{Stixrude2014} for the post-perovskite region. The liquidus is based on a fit that uses the results from \cite{Zerr1998}, \cite{Stixrude2009}, and \cite{Andrault2011}. 
Melt generated at depths shallower than $135$\,km is assumed to erupt instantaneously, meaning that it is transported to the surface, where it solidifies and is represented in the model by solid tracers. This depth is scaled from Earth's neutral buoyancy depth ($\approx 300$\,km) by the ratio of Earth's surface gravity to K2-141\,b's surface gravity. For one model, we also included intrusion, where $70\%$ of the melt that is transported to the surface is instead emplaced at the base of the crust, motivated by estimates from Earth where most mantle derived melt is emplaced intrusively \citep{Crisp1984, White2006}. We assume that eruptions are only possible when the surface is solid. Consequently, no eruptions occur through the magma ocean. Partial melt generated beneath the magma ocean may still be transported to and advected within the magma ocean but is not transported to the surface. 
When melt is produced, trace elements (i.e. water, carbon) are partitioned between solid and molten tracers according to the batch melting model. The concentration of each element in the melt ($C_\mathrm{melt}$) and solid ($C_\mathrm{solid}$) is determined using
\begin{equation}\label{eq:partitioning}
C_\text{melt} = \frac{C_\mathrm{0}}{D+F(1-D)},
\end{equation}
\begin{equation}
C_\text{solid} = D \cdot C_\text{melt},
\end{equation}
where $F$ is the melt fraction, $C_\mathrm{0}$ is the bulk concentration of the cell, and $D$ the partition coefficient of the trace element. 
If the melt is erupted, the trace elements are subsequently outgassed. We use partition coefficients of $D_\mathrm{H}=0.01$ \citep{Aubaud2004}, $D_\mathrm{C}=5.5 \cdot 10^{-4}$ \citep{Rosenthal2015}, following \citet{Liggins2022}. We assume that, upon eruption, these elements are outgassed predominantly as CO$_2$ and H$_2$O, respectively. While the actual speciation depends on oxygen fugacity, we do not explicitly track redox conditions and instead assume that the mantle is sufficiently oxidised to produce CO$_2$ and H$_2$O \citep{lichtenberg2025}. These upper mantle redox conditions might be justified for massive rocky planets with mantles deep enough to self-oxidise via Fe disproportionation \citep{wade_core_2005, Deng2020, hirschmann_magma_2022}. However, it is possible that the mantle of K2-141\,b is more reducing, which would affect some of our results as discussed in section \ref{sec:outgassing}. 

For the initial volatile budget of the mantle, we adopt a present-day bulk-silicate Earth carbon concentration of $110$\,ppm and water concentration of $290$\,ppm \citep{Hirschmann2018}. The volatile budget is assumed to be homogeneously distributed at the start of the simulations. 
At each time step, we compute the surface eruption heat flux as the sum of sensible and latent heat released by the newly erupted material. The sensible heat is calculated from the thermal energy lost as erupted melt cools down from its cell temperature to the temperature of the surface, and is given by: 
\begin{equation}
Q_{\mathrm{sensible}} = \sum_{\mathrm{cell}} m_{\mathrm{eruption}} \cdot c_{\rm p} \cdot (T_\mathrm{pot} - T_{\mathrm{surf}}), 
\end{equation}
where $m_{\mathrm{eruption}}$ is the mass of erupted material per cell, $c_{\rm p}=1200$\,J\,kg$^{-1}$K$^{-1}$ is the specific heat capacity of the melt, $T_{\mathrm{pot}}$ is the adiabatically corrected potential temperature of the melt, and $T_{\mathrm{surf}}$ is the surface temperature. The latent heat contribution is computed as 
\begin{equation}
Q_{\mathrm{latent}} = m_{\mathrm{eruption,tot}} \cdot L,
\end{equation}
where $m_{\mathrm{eruption,tot}}$ is the total amount of erupted material and $L$ is the specific latent heat of melting, for which we use a value of $600$\,kJ\,kg$^{-1}$, representative of silicate rocks \citep{Bea2012}. The heat flux delivered to the surface due to volcanism is then obtained by calculating
\begin{equation} \label{eq:erupt_heatflux}
F_{\mathrm{eruption}} = \frac{Q_{\mathrm{sensible}} + Q_{\mathrm{latent}} }{A}, 
\end{equation}
where $A$ is the surface area of the planet.

To model heat transport in regions of high melt fraction, we parametrise the magma ocean as vigorously convecting fluid with an enhanced effective thermal conductivity $k_\mathrm{h}$, following the eddy diffusivity approach of \cite{Abe1997}. $k_\mathrm{h}$ depends on the local melt fraction $\phi$ and the rheological threshold $\phi_\mathrm{c}\approx0.35$ \citep{Abe1997}, above which the mixture behaves as a low-viscosity fluid. $k_\mathrm{h}$ is given by
\begin{equation} 
k_\mathrm{h} = \exp{\bigg\{\frac{\ln{(k_\mathrm{{h,max}}})}{2}\left[1+\tanh\left({\frac{\phi - \phi_\mathrm{c}}{\Delta\phi}}\right)\right]\bigg\}}-1\,,
\end{equation}
where $\Delta \phi =0.05$ sets the width of the transition and $k_{\mathrm{h,max}} = 10^6$\,Wm$^{-1}$K$^{-1}$ represents the maximum eddy conductivity in the fully molten regime. In this work, we assume that heat transport is only enhanced if the temperature gradient in the magma ocean is super-adiabatic $\big(\frac{dT}{dP} > \frac{dT_\mathrm{a}}{dP}\big)$.
%%%%%%%%%%%%%%%%
%%% FIGURE 1 %%%
%%% END OF FIGURE 1 %%%
%%%%%%%%%%%%%%%%%%%%%%%
\section{Results} \label{sec:results}
%%%%%%%%%%%%%%%%
%%% FIGURE 2 %%%
\begin{figure*}
    \centering
    \includegraphics[width=0.85\textwidth]{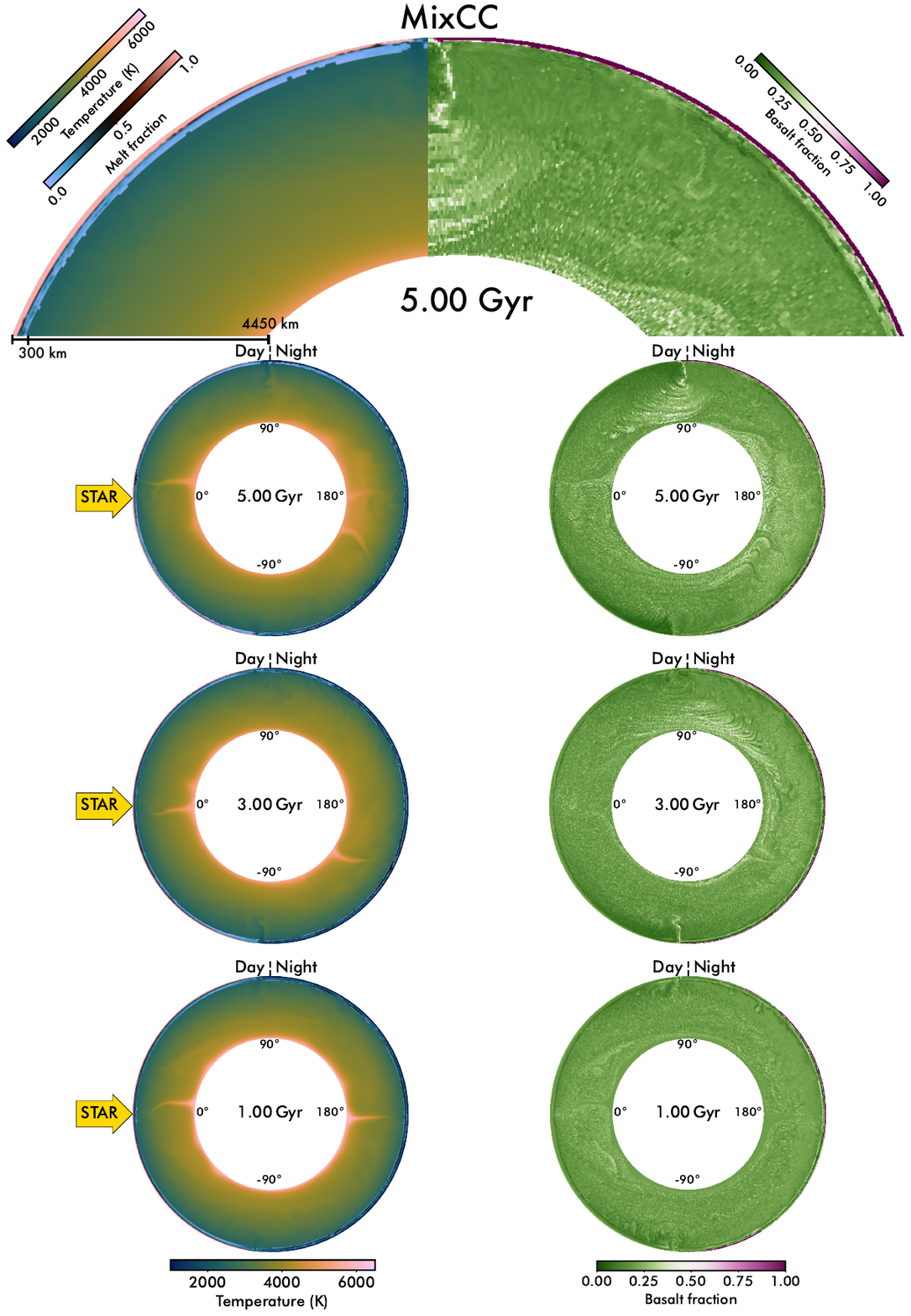}
    \caption{Snapshots of mantle temperature (left column), and basalt fraction (right column) at $1$, $3$, and $5$\,Gyr for MixCC (mixed heating with decreasing CMB temperature and no plastic yielding). The top panel shows a zoom-in of the mantle temperature and basalt fraction at $5\,$Gyr. The model evolves towards a degree-2 convection pattern with stable upwellings near the substellar and antistellar points and downwellings near the day-night terminators. A basaltic crust gradually forms across the nightside, while the dayside remains molten due to sustained high stellar irradiation. The magma ocean is shallow with a thickness around $300$\,km. Videos for this model run are available online and on Zenodo \citep{ZenodoMeier2025}.}
    \label{fig:ref_snapshots}
\end{figure*}
%%% END OF FIGURE 2 %%%
%%%%%%%%%%%%%%%%%%%%%%%
Table~\ref{tab:model_overview} shows an overview of the different models presented in this paper. Models MixCC, MBasal, and MixIntr do not include plastic yielding of the lithosphere (i.e. $\sigma_\text{y}=\infty$\,MPa, strong lithosphere), whereas MPlastic includes plastic yielding with $\sigma_\text{y}=100$\,MPa. MixCC and MPlastic include a decreasing CMB temperature from core cooling. All models except for MBasal (basal heating only) have mixed heating (internal heating + basal heating). MixIntr includes intrusion, with an eruption efficiency of $30\%$ (i.e. $70\%$ of melt is intruded). Throughout this work, we use the term steady state to refer to a quasi-steady convective regime, in which the large-scale flow pattern and thermal structure become approximately time-invariant. In cases where the CMB temperature decreases due to core cooling, a true thermal steady state is not achievable unless sustained by internal heating sources. 
%%%%%%%%%%%%%%%%
%%% TABLE 2 %%%
\begin{table}
\centering
\caption{Overview of model runs}
\label{tab:model_overview}
\begin{tabular}{lcccc}
\hline
\textbf{Model} & $\boldsymbol{\sigma_{y}}$ (MPa) & \textbf{Internal Heating} & \textbf{Core Cooling} & \textbf{Intrusion} \\ 
\hline
MixCC & $\infty$ & mixed heating & \checkmark & $\times$ \\
MBasal & $\infty$ & basal heating & $\times$ & $\times$ \\
MixIntr & $\infty$ & mixed heating & $\times$ & \checkmark\\
MPlastic & $100$ & mixed heating & \checkmark & $\times$ \\
\hline
\end{tabular}
\end{table}
%%%%%%%%%%%%%%%%
%%% END OF TABLE 2 %%%
\subsection{Model MixCC: Mixed Heating with Core Cooling}

As an illustrative case, we first present MixCC using the parameters shown in Table~\ref{tab:parameters}. A steady state is reached after around $3-4$\,Gyr with a mean mantle temperature of approximately $3470$\,K. Figure~\ref{fig:ref_snapshots} shows snapshots of the mantle temperature and the basalt fraction for different time steps. The melt fraction is overlaid on the temperature field using a separate colourbar. During the model run, hot mantle plumes consistently form around the core mantle boundary. The system stabilises in a degree-2 convection pattern, characterised by two prominent superplumes situated at the substellar and antistellar points, and two downwellings at the day-night terminators. If new upwellings form, they eventually merge with the superplume within a few hundred Myrs. 
While this state appears stable, it is important to note that core cooling continues over time, gradually reducing the thermal buoyancy available to drive upwellings from the CMB. As a result, the plumes become weaker (see Fig.~\ref{fig:ref_snapshots}, $5$\,Gyr) and eventually vanish. At $6.5$\,Gyr, the core has reached a temperature of $5100$\,K.
The dayside of the planet is covered by a magma ocean approximately $300$\,km deep ($\approx 3\%\,R_{\mathrm{p}}$), while the nightside surface is solid, with a partially molten layer forming beneath the near-surface.  Within the dayside magma ocean, there is a diverging flow of molten material from the substellar point towards the east and west day-night terminators. On the nightside, a basaltic crust begins to form after approximately $0.6$\,Gyr, covering almost the entire nightside within $1$\,Gyr. The crust and lithosphere, which remain decoupled from the upper mantle flow, undergo intermittent motion towards either the east or west terminator, where it is eventually subducted. The upper mantle, decoupled from the lithosphere, flows from the antistellar point towards the east and west terminators. For most of the time, the lower mantle exhibits a converging flow on both the dayside and nightside, directed towards the substellar and anti-substellar points, respectively. However, when substantial material is subducted, it can occasionally flow past these points, disrupting the typical lower mantle convergence pattern with the flow being more directed towards the substellar point.  

\subsection{Model comparison}
%%%%%%%%%%%%%%%%
%%% FIGURE 3 %%%
\begin{figure}
    \centering
    \includegraphics[width=0.5\textwidth]{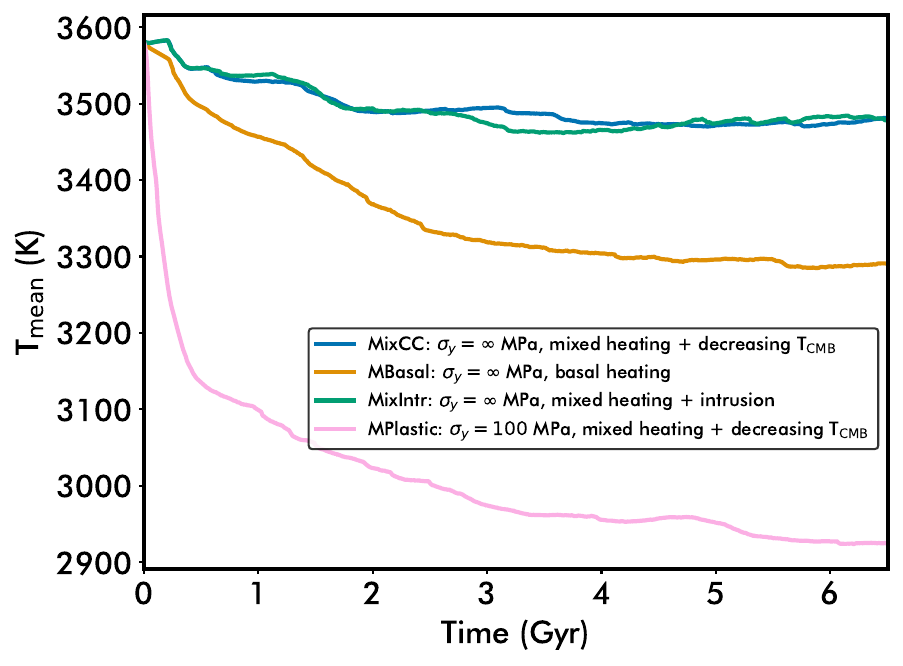}
    \caption{Mean mantle temperature evolution for the different models of K2-141\,b. Models MixCC and MixIntr (both with mixed heating) reach a thermal steady state after approximately $2-3$\,Gyr, with similar final mantle temperatures ($3470$\,K). Model MBasal (basal heating) cools more gradually and reaches a lower steady state temperature ($3300$\,K). Model MPlastic, which includes plastic yielding, exhibits the most significant cooling due to efficient heat loss driven by strong downwellings. Models with decreasing CMB temperature from core cooling (MixCC and MPlastic) are expected to cool further over time, as the core continues to lose heat and thermal buoyancy at the CMB diminishes.}
    \label{fig:tmean}
\end{figure}
%%% END OF FIGURE 3 %%%
%%%%%%%%%%%%%%%%%%%%%%%

Figure~\ref{fig:tmean} shows the mantle temperature evolution for the different models. MixCC and MixIntr, and MBasal reach a steady state after approximately $3-4$\,Gyr. MixCC and MixIntr reach almost the same steady state mean mantle temperature of $3470$\,K, while MBasal settles at a mean mantle temperature of around $3300$\,K. Compared to MixCC, MixIntr includes intrusion and has a constant CMB temperature (effects of core cooling are not taken into account). MPlastic cools the most out of the four models. This is expected, as it is the model that includes plastic yielding, which leads to the formation of more downwellings and therefore a more efficient cooling of the mantle. However, a significant portion of this cooling occurs during the first $500$\,Myr and is therefore mostly dominated by the initial conditions. This model continues to cool beyond several gigayears with a cooling rate of approximately $20$\,K\,Gyr$^{-1}$. At $6.5$\,Gyr, the core of MPlastic has reached a temperature of $5020$\,K. 

%%%%%%%%%%%%%%%%
%%% FIGURE 4 %%%
\begin{figure*}
    \centering
    \includegraphics[width=0.9\textwidth]{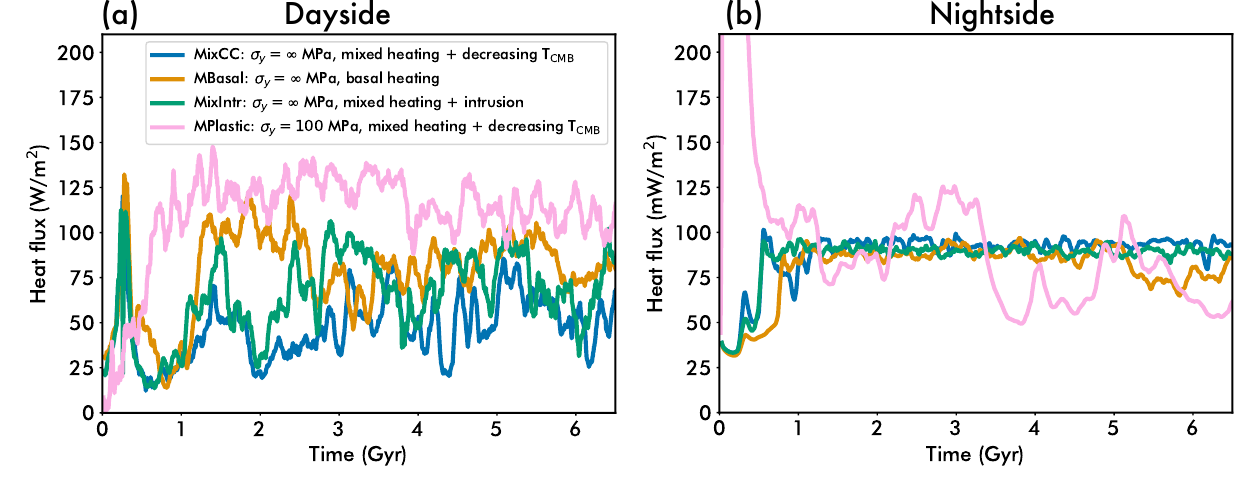}
    \caption{Surface heat flux evolution on the (a) dayside (in W/m$^2$) and (b) nightside (in mW/m$^2$) of K2-141\,b for the different mantle convection models. MixCC, MBasal, and MixIntr show similar trends and stabilise after around $2$\,Gyr, while MPlastic shows higher and more variable heat fluxes due to enhanced downwellings and therefore more efficient mantle cooling.}
    \label{fig:heatflux}
\end{figure*}
%%% END OF FIGURE 4 %%%
%%%%%%%%%%%%%%%%%%%%%%%

Figure \ref{fig:heatflux} shows the averaged dayside (in W\,m$^{-2}$) and nightside (in mW\,m$^{-2}$) outgoing heat flux for the different models (excluding heat flux from eruptions). For MPlastic, there is a high nightside heat flux because of the initial formation of downwellings which leads to a large decrease in mantle temperature (Fig.~\ref{fig:tmean}). As expected from their final mantle temperatures, MixCC and MixIntr also have similar surface nightside heat fluxes. The nightside heat flux for MBasal is similar to that of MixCC and MixIntr, but it has a higher dayside heat flux for approximately $2$\,Gyrs between $1$ and $3$\,Gyr. Together with the absence of internal heating, this leads to an overall cooler mantle temperature compared to MixCC and MixIntr. For MixIntr, a large increase in nightside heat flux is observed at around $4.1$\,Gyr, rising from approximately $50$\,mW\,m$^{-2}$ to over $120$\,mW\,m$^{-2}$, which is most likely due to the merging of plumes. 

%%%%%%%%%%%%%%%%
%%% FIGURE 5 %%%
\begin{figure*}
    \centering
    \includegraphics[width=0.9\textwidth]{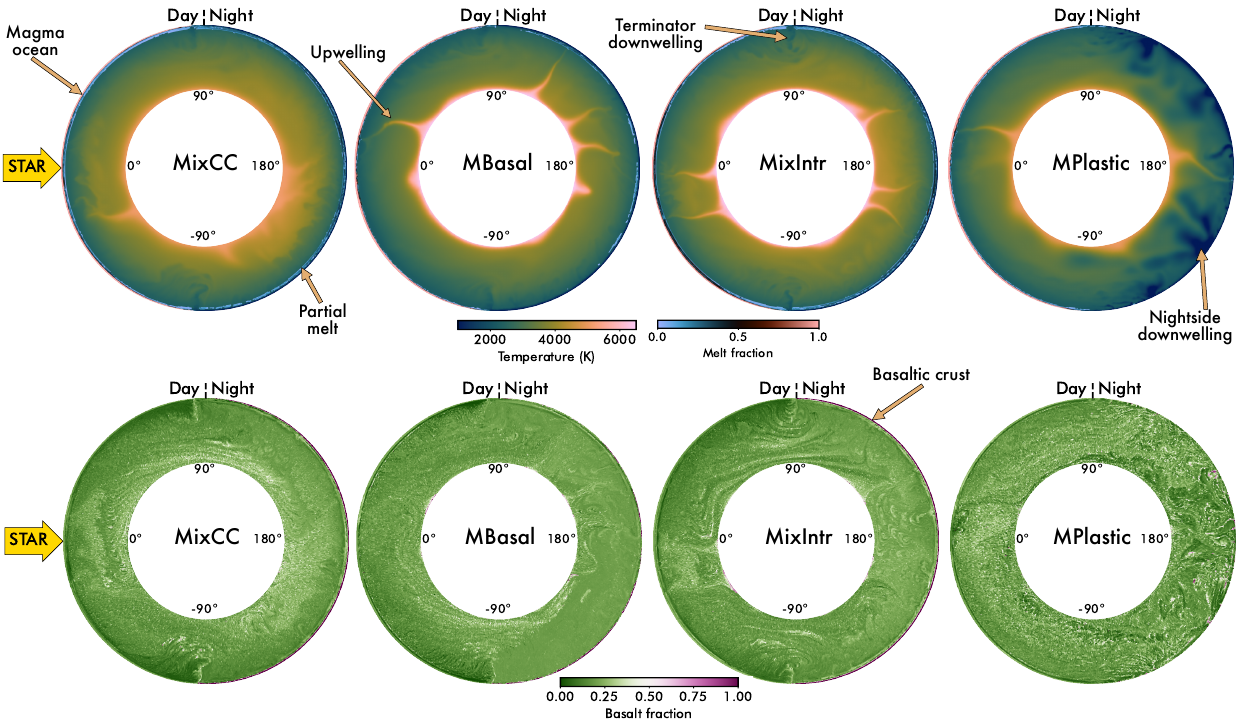}
    \caption{Snapshots of mantle temperature and melt fraction (top row) and basalt fraction (bottom row) at $6.5$\,Gyr for each interior model of K2-141\,b. All models show a shallow magma ocean on the dayside (left hemisphere) and a solid nightside (right hemisphere). Models without plastic yielding (MixCC, MBasal, MixIntr), exhibit upwellings at the substellar and antistellar points, with downwellings forming at the day-night terminators. MPlastic, which includes plastic yielding (lithospheric weakening), displays a more asymmetric pattern, with downwellings forming on the nightside and upwellings rising preferentially on the dayside. Videos for these model runs are available online and on Zenodo \citep{ZenodoMeier2025}.}
    \label{fig:5gyrplot}
\end{figure*}
%%% END OF FIGURE 5 %%%
%%%%%%%%%%%%%%%%%%%%%%%

Figure \ref{fig:5gyrplot} shows snapshots of the mantle temperature (top row) and basalt fraction (bottom row) for the different models after 6.5\,Gyr of evolution. MBasal and MixIntr exhibit similar behaviour to MixCC in terms of downwelling and crustal formation: For these models, a basaltic crust forms on the nightside with a partially molten layer beneath the near-surface. 
The crustal thickness for MixCC and MixIntr is around $80$ and $100$\,km, respectively. The crustal thickness of MBasal shows more variation over time: After an initial rapid growth phase during the first $1$\,Gyr, the crustal thickness gradually increases from about $40$\,km to $80$\,km over the next $3$\,Gyr. This is followed by a phase of thinning, with the crust decreasing to around $42$\,km over the subsequent $3$\,Gyr. Towards the end of the simulation, the crust begins to thicken again.   
Two downwellings form near the day-night terminators with crustal material being subducted either at the eastern or western terminator. One can see that MBasal and MixIntr are characterised by more numerous and stronger upwellings compared to MixCC and MPlastic. This is because MBasal and MixIntr assume a constant CMB temperature and therefore do not account for core cooling, which in the other models leads to a decreasing CMB temperature over time and reduced buoyancy at the base of the mantle.  On the other hand, MPlastic, which includes plastic yielding, is characterised by several downwellings on the nightside, while upwellings develop around the CMB and tend to migrate preferentially towards the dayside. If an upwelling on the nightside reaches the surface, it can lead to melt production. There is no partially molten layer below the near-surface in MPlastic and no pronounced downwellings form near the terminator regions in this model. For the first $500$\,Myr, the (partially molten) magma ocean extends to approximately $30\degree$ into the nightside. As the mantle mean temperature cools to around $3150$\,K (at around $500$\,Myr; see Figure~\ref{fig:tmean}), the magma ocean retreats and its boundary stabilises near the terminators. However, lateral flow continues to partially melt the upper mantle in this region. This leads to the formation of numerous small-scale downwellings, which are subsequently advected back towards the dayside.

\begin{figure*}
    \centering
    \includegraphics[width=0.9\textwidth]{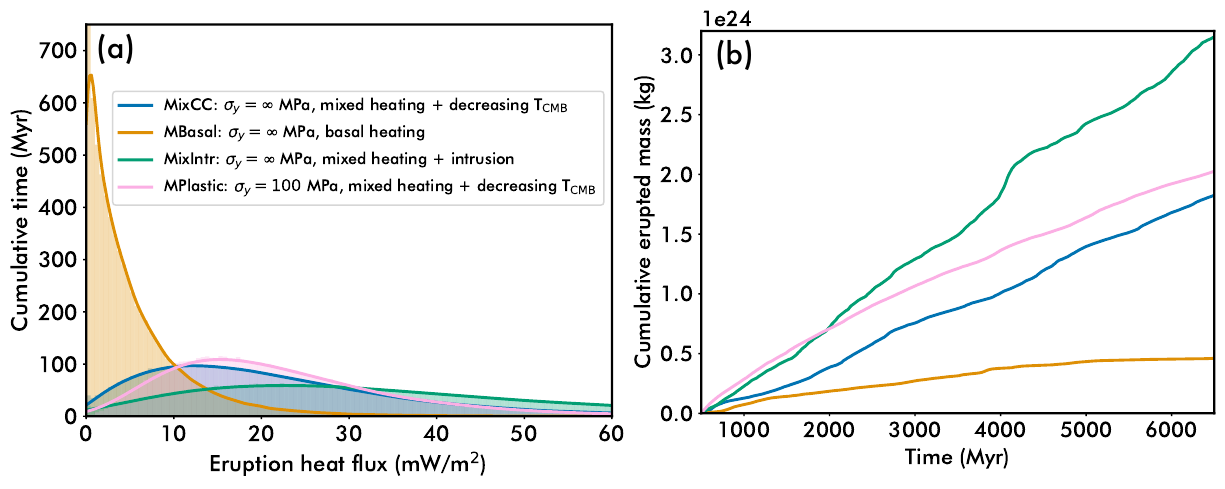}
    \caption{(a) Time-weighted histograms of eruption heat flux for the four interior dynamic models of K2-141\,b. Each curve shows the cumulative time spent within a given eruption heat flux range, weighted by the duration of each model time step. MBasal is dominated by low-flux, long-lived phases, whereas MixCC, MixIntr and MPlastic spend more time in higher flux ranges. (b) Cumulative erupted lava mass as a function of time. All models show continuous volcanic activity, but with distinct eruption histories: MixIntr leads to the largest total erupted mass, while MBasal produces less due to the lack of internal heating and therefore more limited melting.}
    \label{fig:eruption_plot}
\end{figure*}

Figure \ref{fig:eruption_plot}~a shows the time-weighted histograms of the surface eruption heat flux for the different models, calculated using Equation (\ref{eq:erupt_heatflux}). Each histogram bin reflects the cumulative time the planet spends experiencing a given range of eruption heat flux values, weighted by the model time step duration. We excluded the first $500$\,Myr, as these are dominated by the initial condition and spin-up time of the model leading to an increased eruption rate.  
The basally heated model (MBasal) is dominated by low eruption heat flux (a few mW/m$^2$), while models with mixed heating (MixCC, MixIntr, MPlastic) exhibit broader distributions that extend to higher heat fluxes on the order of $10-100$\,mW/m$^2$. However, even the highest eruptive heat fluxes remain lower than the conductive heat flux through the nightside (Fig.~\ref{fig:heatflux}~b), so that eruptions play a negligible role in warming up the nightside (see also Section~\ref{sec:discuss_nightside} for further discussion of nightside heat flux).  
Figure \ref{fig:eruption_plot}~b shows the cumulative erupted mass as a function of time, where again the first $500$\,Myr have been excluded. MixIntr leads to the largest total erupted mass, whereas MBasal produces the least amount of erupta, which is most likely due to the lack of internal heating and therefore more limited melting. Although MixIntr exhibits lower erupted mass than MPlastic during the first $2$\,Gyr, it eventually surpasses MPlastic, leading to a higher cumulative erupted mass over geological timescales. The cumulative mass for MixIntr and MixCC increases nearly linearly over time, suggesting a steady eruption rate. In contrast, MBasal and MPlastic display a decelerating trend, with eruption rates that decline over the course of the simulation. 

\subsection{Surface composition}
\begin{figure*}
    \centering
    \includegraphics[width=0.9\textwidth]{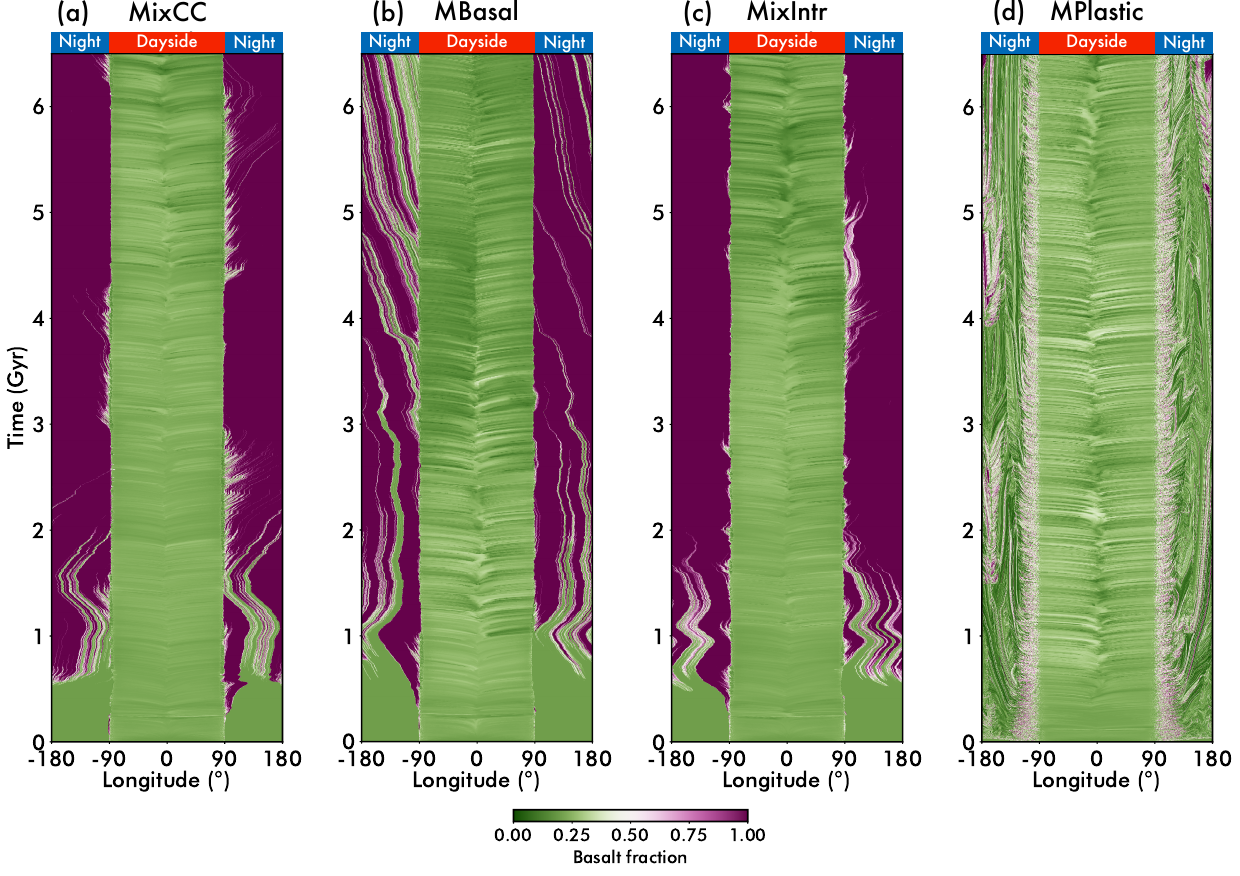}
    \caption{Time evolution of the surface basalt fraction for the different interior models of K2-141\,b. Each panel shows the basaltic content (0-1) across longitude and time from $0$ to $5$\,Gyr: (a) MixCC (mixed heating with decreasing CMB temperature from core cooling), (b) MBasal (basal heating only), (c) MixIntr (mixed heating with intrusion), (d) MPlastic (mixed heating with plastic yielding and decreasing CMB temperature from core cooling). The dayside spans $-90\degree$ to $+90\degree$. For MixCC, MBasal, and MixIntr, a basaltic crust forms on the nightside after around $0.5$\,Gyr and is gradually recycled at the day-night terminators. MPlastic shows a more varied surface composition with frequent recycling of basalt due to widespread downwellings.}
    \label{fig:composition_plot}
\end{figure*}

Figure \ref{fig:composition_plot} shows the evolution of the surface composition in terms of the total (solid and melt) basalt fraction. All models start with a basalt fraction of $0.2$ and a harzburgite fraction of $0.8$. For all models, the magma ocean generally keeps a basalt fraction around $0.2$. There is a lot of variation on short time scales where the surface can be locally depleted (basalt fraction near zero) or slightly more enriched (around $0.5$). These fluctuations are caused by high velocities within the magma ocean, where diverging flow from the substellar point rapidly redistributes melt towards the nightside. Subsequently, the magma ocean is replenished with fresh mantle material from the upper mantle, which has a basalt fraction around $0.2$. We note, however, that the velocities in the magma ocean are not physically accurate, as the viscosity in the region remains too high to capture the true dynamics of the magma ocean. This is due to the lower viscosity cutoff of $10^{16}$\,Pa\,s imposed to ensure numerical stability of the model. Flow speeds in the magma ocean reach $10-100$\,cm\,yr$^{-1}$, compared to a few cm\,yr$^{-1}$ in the solid mantle. 
On the nightside, a basaltic crust starts to form after the initial spin-up phase of around $500$\,Myr for models MixCC, MBasal, and MixIntr. After $2$\,Gyr, MixCC and MixIntr have a surface that is fully basaltic. MBasal (basally heated) also develops a predominantly basaltic surface, but retains a few regions with lower basalt content. 
For all models with no plastic yielding (i.e. $\sigma_y=\infty$, models MixCC, MBasal, and MixIntr), two pronounced downwellings form at the eastern and western day-night terminators at the interface between the hot magma ocean and the cold, solid nightside. Despite the symmetric placement of these downwellings, the surface crust moves as a single lid across the nightside, resulting in crustal recycling at only one terminator. For MixCC and MixIntr, the lid exhibits intermittent eastward and westward motion, which indicates a decoupling between the crust and the underlying upper mantle, which exhibits a diverging flow from the antistellar point towards the eastern and western terminator. For MBasal, the lid undergoes fewer changes in direction and material is preferentially recycled at one terminator. For MPlastic, which includes a weaker lithosphere with $\sigma_y = 100$\,MPa, no such single-lid basaltic crust forms on the nightside because downwellings continuously subduct material with higher basaltic content back into the mantle.      
%%%%%%%%%%%%%%%%
%%% FIGURE 6 %%%
%%% END OF FIGURE 6 %%%
%%%%%%%%%%%%%%%%%%%%%%%

%%%%%%%%%%%%%%%%
%%% FIGURE 7 %%%
%%% END OF FIGURE 7 %%%
%%%%%%%%%%%%%%%%%%%%%%

\section{Discussion}  \label{sec:discussion}

\subsection{Convection regimes}
Our models reveal a range of mantle convection regimes depending on the interior heating mode and strength of the lithosphere. For the models with no plastic yielding ($\sigma_{y}=\infty$), the mantle is characterised by a degree-2 convection pattern with upwelling regions near the substellar and anti-substellar points and downwellings forming near the day-side terminators. Below the lithosphere, a diverging flow is established with material flowing from the antistellar point towards the dayside. Occasionally, this can drive the flow upwellings towards the dayside. However, these downwellings are not able to cross the day-night terminator because the highly viscous downwelling flow counters the motion of upwellings back towards the nightside. The formation of such downwellings is a direct consequence of the hemispheric temperature contrast with a magma ocean on the dayside and a cold, non-molten nightside. 

It is worth pointing out that these terminator downwellings form despite not using a plastic yielding criterion for the strength of the lithosphere, which is atypical for stagnant lid convection \citep[e.g.][]{Tackley2000, Lourenco2023}. The hot, fully molten dayside surface and the cold, solid nightside surface (for MixCC, MBasal, MixIntr) establish strong lateral thermal gradients across the terminator. These gradients drive a flow that induces enough stress at the terminators to destabilise the lithosphere. On Earth, subduction occurs primarily because the oceanic crust is moving away from the diverging mid-ocean ridges, cools down and becomes dense enough to sink into the mantle \citep[e.g.][]{Schubert2001, Mulyukova2020}. This process is further sustained by large-scale plate motions, slab pull \citep[e.g.][]{Forsyth1975} and hydration of the lithosphere \citep[e.g.][]{RegenauerLieb2001}, where water incorporated into the oceanic plate weakens it and helps initiate and sustain subduction. These processes are not included in our models and differ from those responsible for the formation of the modelled terminator downwellings in this study. We attribute the formation of these downwellings to the convergence of mantle flow at the day-night boundary. On the dayside, the flow is diverging, driven by the temperature contrast. On the nightside, mantle material beneath the lithosphere also flows from the antistellar point towards the terminators. These two flows (dayside/nightside) are disconnected but converge laterally at the day-night terminators. Because the lithosphere is cold and relatively stiff and the surface boundary condition is free-slip, the converging horizontal flow at the terminator cannot be accommodated by surface deformation. To satisfy mass conservation under these constraints, built-up stresses are relieved by downward deflection of the flow, resulting in localised terminator downwellings. It is also worth noting that this tectonic regime is not plate tectonic-like since the lithosphere is not divided into several strong plates that move with respect to each other \citep{Stern2023}. In fact, the planet possesses only a single `plate': the dayside is entirely molten and thus lacks a lithosphere, rendering the planet a single-lid system. The novelty of this regime lies in the fact that recycling at the terminators can occur despite the presence of only one plate, enabled by the hemispheric magma ocean. 

However, it is important to consider how the two-dimensional geometry used in this study could influence the observed convection patterns and terminator downwellings. In our case, the 2D spherical annulus setup favours large-scale, degree-2 convection modes. In a fully three-dimensional spherical flow, the flow can become more time-dependent and spatially variable \citep[e.g.][]{Schubert2001, Travis1990}. Such time-dependent behaviour could give rise to multiple or laterally migrating downwellings that may disturb the symmetry and long-term stability of the terminator-recycling regime identified here. Moreover, our simulations cannot capture the sheet-like geometry of downwelling slabs that develop in three-dimensional convection \citep{Bercovici1989}, where downwellings would occur as long, linear structures. Future work should therefore assess how the inclusion of 3D geometry affects the convection pattern and stability of terminator downwellings.

\begin{figure}
    \centering
    \includegraphics[width=0.5\textwidth]{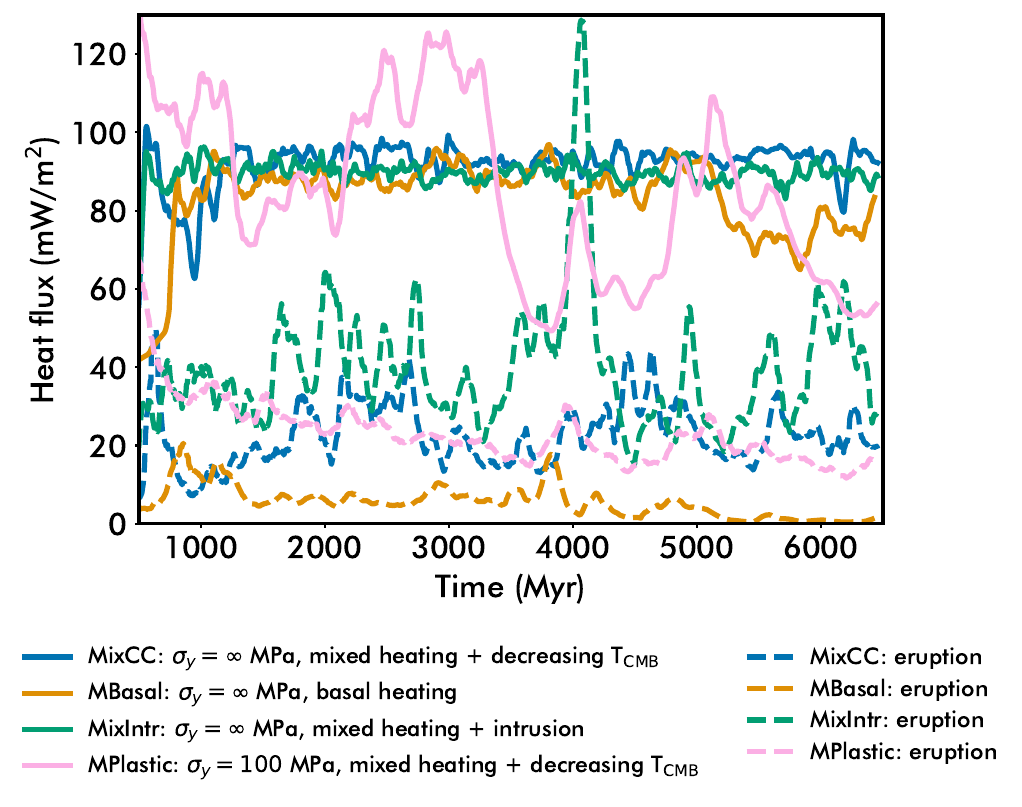}
    \caption{Averaged nightside surface heat flux evolution for the different mantle convection models of K2-141\,b. Solid lines show the conductive heat flux. Dashed lines show the heat flux from surface eruptions (latent + sensible heat). Models with mixed heating (MixCC, MixIntr, MPlastic) show higher and more variable volcanic heat flux compared to the basally heated model (MBasal).}
    \label{fig:eruption_conduction_plot}
\end{figure}

The nightside is characterised by efficient cooling, facilitated by downwelling flow at the terminators and conductive heat loss through the lithosphere. Figure~\ref{fig:eruption_conduction_plot} shows both the conductive heat loss (solid lines) and the heat loss from eruptions (dashed lines) for the different models. Although volcanic eruptions contribute to heat loss, on average it is lower than the conductive heat flux. This indicates that the tectonic mode on the nightside is not heat-pipe like \citep{Moore2001, Tackley2023}, despite the high eruption efficiency. A heat-pipe mode would likely require higher internal heating rates, as suggested for the early Earth and Io \citep{Moore2013, Moore2017}. In this study, we adopt present-day Earth-like internal heating rates. Future work could explore the implications of higher internal heating, as might be expected for a younger or more tidally heated planet. 
In the following section, we discuss whether such a nightside heat flux could be observed with telescopes such as JWST. 
%%%%%%%%%%%%%%%%
%%% FIGURE 8 %%%
%%% END OF FIGURE 8 %%%
%%%%%%%%%%%%%%%%%%%%%%%

\subsection{Observability of nightside heat flux} \label{sec:discuss_nightside}
Spitzer observations of K2-141\,b indicate that the planet most likely lacks a thick atmosphere and has a nightside temperature that is consistent with zero kelvin \citep{Zieba2022}. Figure~\ref{fig:eruption_conduction_plot} shows the expected heat flux on the nightside from the cooling of the mantle as well as the heat flux expected from volcanic eruptions. In total this would lead to a heat flux on the order of around $120$-$150$\,mW/m$^2$, which would only lead to a temperature increase of around $40$\,K (assuming a $0$\,K nightside surface with unit emissivity). Sensitivity to this small heat flux is beyond the capabilities of current space telescopes such as JWST. 

The heat flux from volcanic eruptions needs to be analysed a bit more carefully, however. The heat flux shown in Figure~\ref{fig:eruption_conduction_plot} is the average heat flux over the duration that simulation snapshots were taken, which corresponds to $100\cdot dt$, where $dt$ is the model time step. The heat flux calculated at each time step is highly variable, with periods where it is close to zero (no eruptions) and periods where the heat flux can reach an intensity on the order of W/m$^2$. However, this heat flux is still undersampled, limited by the time step duration of the model, being on the order of tens of thousands of years. The heat flux output from the model therefore corresponds to the total heat flux over this duration. In reality, volcanic eruptions occur on much shorter time scales ($\mathcal{O}$(days)). 
We can make a simple back-of-the-envelope calculation to estimate the eruption heat flux for shorter eruption time scales: Let $F_\text{avg}$ be the average heat flux over the duration of a model time step $\Delta t$.  We also note $f_\text{erupt}$ the eruption frequency, i.e. how many eruptions occur per day, $\tau$ the duration of an eruption, and $A_\text{erupt}$ the area over which a typical eruption occurs. The number of eruptions $N_\text{erupt}$ is therefore given by 
\begin{equation}
N_\text{erupt} = \Delta t \cdot f_\text{erupt}.
\end{equation}
The energy per eruption can be calculated as
\begin{equation}
E_\text{erupt} = \frac{F_\text{avg} A_\text{planet} \Delta t}{\Delta t f_\text{erupt}},
\end{equation}
where $A_\text{planet}$ is the surface area of the planet. Subsequently the eruption heat flux is given by
\begin{equation}
F_\text{erupt} = \frac{E_\text{erupt}}{A_\text{erupt}\tau} = \frac{F_\text{avg}A_\text{planet}}{A_\text{erupt}\tau f_\text{erupt}}
\end{equation}
From this equation, we can identify two factors that will amplify the heat flux: $f_{\rm A} = \frac{A_\text{erupt}}{A_\text{planet}}$ is the fraction of the surface that is erupting. On the other hand, $D = \tau f_\text{erupt}$ quantifies how continuous or sporadic eruptions occur. This can be thought of analogous to electronics, where the Duty cycle (or Duty factor) represents the fraction of one period in which a signal is active. $D=1$ corresponds to a continuous eruption and $D \ll 1$ corresponds to eruptions that occur rarely and over a short period of time (sporadic bursts of volcanic eruptions). In the following, we assume that $f_{\rm A}$ and $D$ follow a Log-normal distribution, i.e. $\ln{f_{\rm A}} \sim \mathcal{N}(\mu_{f_{\rm A}},\sigma_{f_{\rm A}}^2)$ and $\ln{D} \sim \mathcal{N}(\mu_{\rm D},\sigma_{\rm D}^2)$. For simplicity, this choice is motivated by Earth-based studies suggesting that several volcanic parameters follow log-normal statistics \citep[e.g.][]{Sanchez2012}. However, the magnitude of eruptions---which we assume here to be analogous to the erupting fraction $f_{\rm A}$---could also follow different statistical distributions, such as a Poisson distribution \citep[e.g.][]{Marzocchi2006, Deligne2010}. If we further assume that $f_{\rm A}$ and $D$ are uncorrelated (although this is probably not the case in reality), the mean erupted flux is then given by
\begin{equation}
\langle F_\text{erupt} \rangle = F_\text{avg} \langle f_{\rm A}^{-1} \rangle \langle D^{-1} \rangle,
\end{equation}
with
\begin{equation}
\langle f_{\rm A}^{-1} \rangle = \int_0^1 \frac{p(f_{\rm A})}{f_{\rm A}}df_{\rm A} = \exp{(-\mu_{f_{\rm A}} +\frac{1}{2}\sigma_{f_{\rm A}}^2)},
\end{equation}
and
\begin{equation}
\langle D^{-1} \rangle = \int_0^1 \frac{p(D)}{D}dD = \exp{(-\mu_{\rm D} +\frac{1}{2}\sigma_{\rm D}^2)},
\end{equation}
where $p(f_{\rm A})$ and $p(D)$ are the Log-normal distributions of $f_{\rm A}$ and $D$.
We can now calculate the planet-to-star contrast using
\begin{equation}
\left(\frac{F_{\rm p}}{F_{\rm *}} \right) = f_A \cdot \left(\frac{R_{\rm p}}{R_{\rm *}}\right)^2 \cdot \frac{\int_{\lambda_{\text{min}}}^{\lambda_\text{max}} B_{\mathrm{\lambda}}(T_\text{erupt})}{\int_{\lambda_{\text{min}}}^{\lambda_\text{max}} B_{\mathrm{\lambda}}(T_{\rm *})},
\end{equation}
where $R_{\rm p}$ is the radius of the planet, $R_{\rm *}$ the radius of the star, and $B_{\rm \lambda}(T)$ is the Planck function. $\lambda_{\text{min}}$ and $\lambda_{\text{max}}$ correspond to the wavelength coverage of the observations. We assume that the erupting surface radiates as a blackbody and the brightness temperature $T_\text{erupt}$ is therefore given by
\begin{equation}
    T_\text{erupt} = \left(\frac{\langle F_\text{erupt}\rangle}{\sigma}\right)^{1/4},
\end{equation}
%%%%%%%%%%%%%%%%
where $\sigma=5.67\cdot10^{-8}$\,Wm$^{-2}$K$^{-4}$ is the Stefan-Boltzmann constant. 
%%% FIGURE 9 %%%
\begin{figure}
    \centering
    \includegraphics[width=0.5\textwidth]{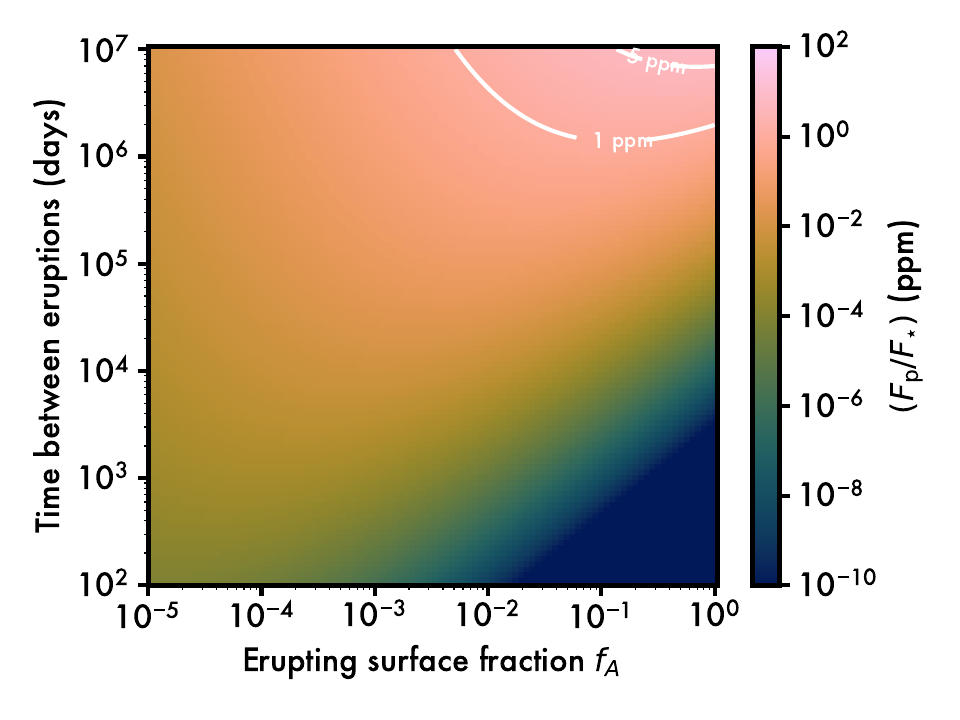}
    \caption{Estimated planet-to-star flux contrast $F_{\rm p}/F_{\rm *}$ in the MIRI wavelength range ($10-15\,\mu$\,m) on the nightside of K2-141\,b, resulting from volcanic eruptions. The colour scale shows the flux ratio in ppm as a function of the erupting surface fraction $f_{\rm A}$ and the time between eruptions (assuming a fixed eruption duration of $\tau = 10$\,days). The contours indicate $1$ and $5$ppm. These estimates assume black body emission and log-normal distribution for $f_{\rm A}$ and the eruption duty cycle $D$.}
    \label{fig:eruption_miri_ppm}
\end{figure}
%%% END OF FIGURE 9 %%%
%%%%%%%%%%%%%%%%%%%%%%%
Figure \ref{fig:eruption_miri_ppm} shows the corresponding planet-to-star flux in the JWST/MIRI wavelength range, as a function of the erupting surface fraction $f_A$ and the time between eruptions. We determine the latter from $1/f_\text{erupt} = \frac{\tau}{D}$, assuming an average eruption duration of $\tau=10$\,days. 
For this figure, we assumed that $F_\text{avg}=0.1$\,W/m$^{2}$. Contours indicate flux ratios of $1$ and $5$ ppm. 
If eruptions occur frequently (i.e. shorter time between eruptions), the average heat flux $F_\text{avg}$ is spread over many events, resulting in relatively cooler individual eruptions and a lower planet-to-star flux ratio. Conversely, if eruptions are rare (i.e. longer time between eruptions), the same $F_\text{avg}$ must be emitted during fewer, shorter episodes, requiring each eruption to be hotter and more intense, thereby increasing the thermal contrast. 
For instance, if $10\%$ of the planet's surface is erupting and such eruptions occur roughly every $800{,}000$\,days (2192 years), the resulting flux contrast would be $\approx 1$\,ppm. This is significantly below the estimated noise floor limit of MIRI LRS of $\approx 25$\,ppm \citep[e.g.][]{Bouwman2023}, even when multiple eclipses are observed. For instance, \texttt{PandExo} \citep{Batalha2017} simulations of JWST observations with MIRI LRS or NIRSpec show that in order to distinguish a difference of approximately $50$\,K in temperature requires at least 189 eclipses in the 10-15 $\mu$m bandpass after binning to 0.44 $\mu$m per bin. Assuming a baseline as long as the eclipse duration of 0.94 hours \citep{Barragan2018}, the program would require approximately 355 hours of science with JWST.
This analysis demonstrates that modulations in the thermal phase curve due to volcanic activity are extremely challenging to detect with current or near-future space-based observatories, especially since the more detectable, high-intensity eruptions must occur infrequently by construction, reducing the likelihood of observing one during a single eclipse or phase curve observation.  

While thermal emission from nightside volcanism is unlikely to be detectable, the spatial distribution of volcanic activity could have important consequences for where outgassing could be detected on an exoplanet. 
%%%%%%%%%%%%%%%%
%%% FIGURE 10 %%%
\begin{figure*}
    \centering
    \includegraphics[width=0.9\textwidth]{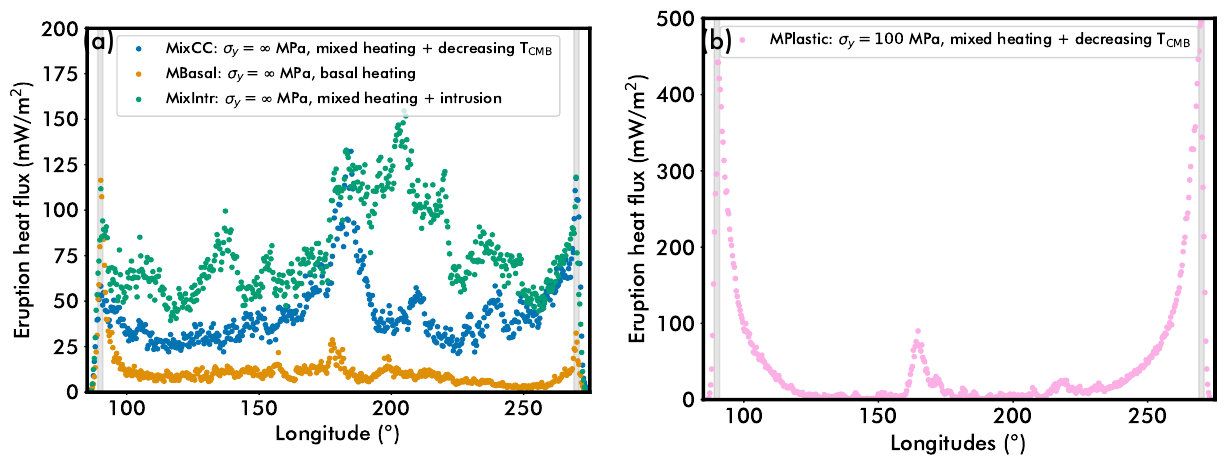}
    \caption{Longitudinal distribution of eruption heat flux on the nightside of K2-141\,b, averaged over time. (a) Models without plastic yielding. (b) Model with plastic yielding. The shaded vertical bands denote the day-night terminator regions. In MPlastic, volcanic activity is concentrated near the day-night boundary, while MixCC, MBasal, and MixIntr exhibit more widespread eruptions across the nightside with some increase near the terminators. The spatial distribution could influence where outgassed volatiles appear in transmission spectra.}
    \label{fig:eruption_longitude}
\end{figure*}
%%% END OF FIGURE 10 %%%
%%%%%%%%%%%%%%%%%%%%%%%
To explore the spatial variability of volcanic activity, we compute the longitudinally averaged eruption heat flux for each model by averaging the 2D eruption heat flux field over time. This yields the mean eruption heat flux per longitude bin.  
Figure~\ref{fig:eruption_longitude} shows the resulting longitudinal distribution of eruption heat flux for models MixCC, MBasal, MixIntr (panel a) and MPlastic (panel b). The shaded regions indicate the day-night terminator regions. For MPlastic, we again excluded the first $500$\,Myr to remove the initial model spin-up. In some cases, outgassing is concentrated near the day-night terminators. This raises the possibility that volatile species outgassed by volcanism could imprint detectable signatures in transmission spectra, even if the total heat flux remains too low to affect the thermal phase curve. However, this depends on the chemical lifetimes and atmospheric mixing of the outgassed species. Future studies incorporating coupled interior-atmosphere models will be essential to assess whether terminator volcanism could lead to observable signatures. 

\subsection{Magma ocean thickness and dayside heat flux}
%%%%%%%%%%%%%%%%
%%% FIGURE 11 %%%
\begin{figure}
    \centering
    \includegraphics[width=0.5\textwidth]{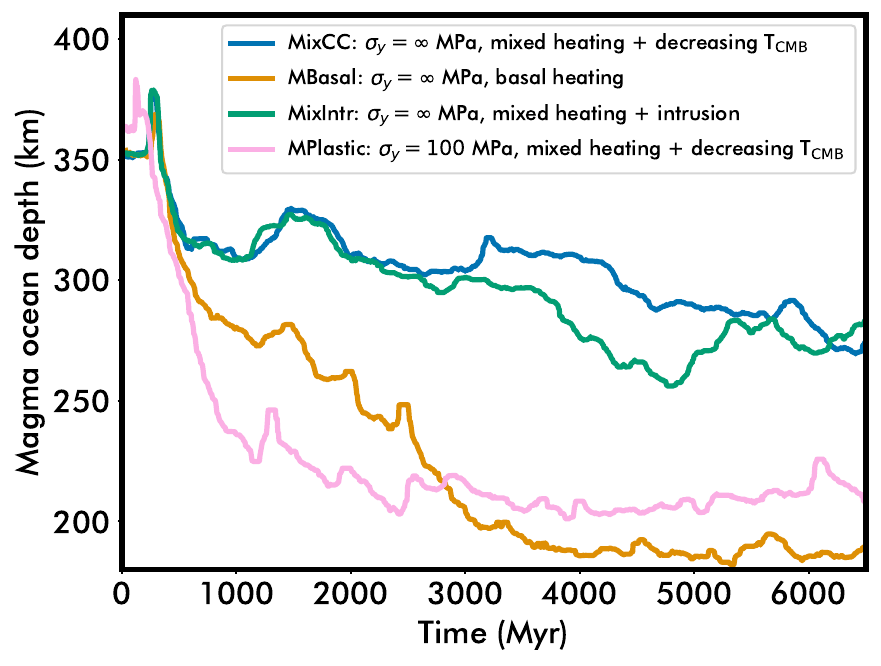}
    \caption{Evolution of the dayside magma ocean thickness for the different models of K2-141\,b. The magma ocean thickness is laterally averaged across the molten dayside. All models exhibit an initial rapid thickening due to early impinging plumes, followed by gradual thinning as the mantle cools. MixCC and MixIntr maintain a thicker magma ocean ($\sim 280-310$\,km), while MBasal (basal heating) and MPlastic (with plastic yielding) result in shallower magma oceans ($\sim 200$\,km).}
    \label{fig:mo_thickness}
\end{figure}
%%% END OF FIGURE 11 %%%
%%%%%%%%%%%%%%%%%%%%%%%
\cite{Meier2023} showed that the magma ocean thickness on lava worlds such as 55 Cancri e is only a small fraction of the whole mantle depth, in the absence of a thick atmosphere, which is a consequence of the sub-adiabatic temperature profile in the magma ocean. This has been further demonstrated by \cite{Lai2024}, who showed that for negligible internal heat sources the magma ocean thickness remains on the order of hundreds of meters. Figure \ref{fig:mo_thickness} shows the evolution of the magma ocean thickness for the different models of K2-141\,b. The thickness corresponds to the laterally averaged thickness of the dayside. MixCC and MixIntr reach a magma ocean thickness of approximately $280$ and $310$\,km, whereas MBasal and MPlastic have a shallower thickness of around $200$\,km. This result is consistent with \cite{Boukare2025}, who conducted 2D and 3D magma ocean dynamics simulations for K2-141\,b and found that, once the planet enters its solid-state mantle stage, the dayside magma ocean remains shallow with a thickness of less than $200$\,km. Compared to the study of \cite{Lai2024}, our models produce deeper magma oceans because we include internal heat sources and our magma ocean parametrisation does not capture the overturning circulation, which could lead to shallower magma oceans \citep{Kite2016, Lai2024}. Additionally, the limited vertical resolution in our models (64 grid cells) may affect the precise determination of the magma ocean depth. A future study could explore the sensitivity of the magma ocean depth to vertical resolution by running higher-resolution models. 

All models in this work show a sudden increase in the magma ocean thickness within the first few hundred Myrs, which is driven by the initial condition with several hot upwellings rising towards the dayside. For MBasal and MPlastic, the mantle takes longer to reach a steady state in terms of mean mantle temperature (Fig.~\ref{fig:tmean}) compared to MixCC and MixIntr and the magma ocean thickness therefore also takes longer to decrease for these models. At around $1.2$\,Gyr, the magma ocean thickness increases by roughly $25$\,km for MPlastic. This is because several plumes that are rising from the CMB merge together and subsequently heat up the magma ocean leading to increased melt production. A similar event happens at $2.5$\,Gyr for both MPlastic and MBasal as well as at $1.2$\,Gyr for MixCC and MixIntr and at $3.2$\,Gyr for MixCC. Models with decreasing CMB temperature are expected to have fewer of these events over time since plume formation is driven by the thickness of the lower thermal boundary layer. The heat flux on the dayside also increases following such impinging plume events. From Figure \ref{fig:heatflux} a one can see that the flux on the dayside of MixIntr increases to around $100$\,W/m$^2$ and then drops again to around $25$\,W/m$^2$ within $500$\,Myr. It then increases again at $2$\,Gyr following the merging with another plume, although this does not lead to a much thicker magma ocean in that case. Using the Stefan--Boltzmann law, an increased internal heat flux of $100$\,W/m$^2$ would raise the dayside temperature by only a few kelvin. This is because the dayside radiative flux is already very high ($\approx 90\,000$ W/m$^2$ for an equilibrium temperature of $2000$\,K). Such a modest increase contributes only a fraction of a ppm to the secondary eclipse depth, making it undetectable with current or near-future telescopes.

\subsection{Volcanic outgassing}\label{sec:outgassing}
Continuous eruptions on the nightside lead to the outgassing of volatile species. 
%%%%%%%%%%%%%%%%
%%% FIGURE 12 %%%
\begin{figure}
    \centering
    \includegraphics[width=0.5\textwidth]{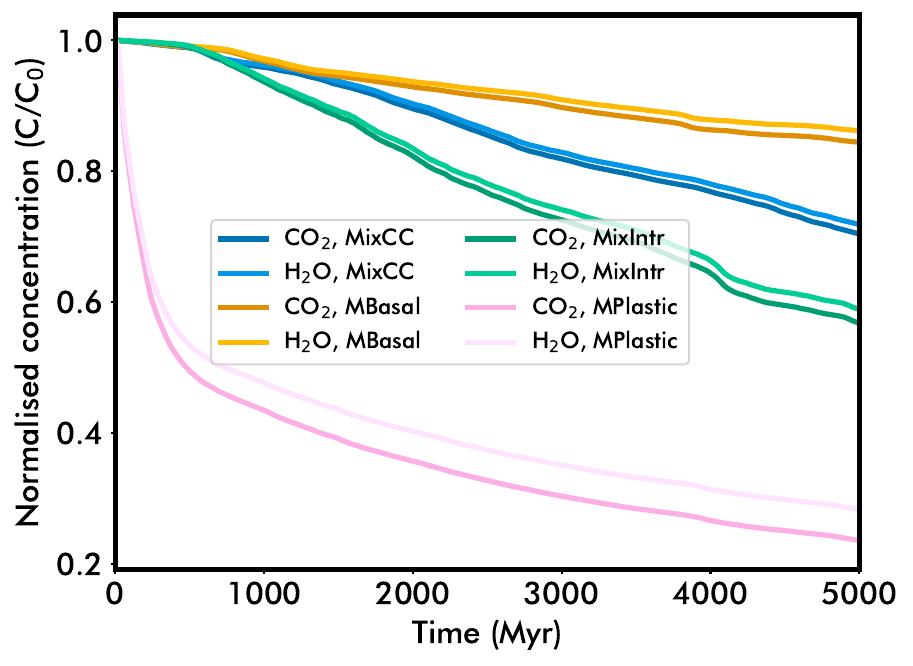}
    \caption{Normalised volatile concentration in the mantle of K2-141\,b for the different models. Volatiles (CO$_2$, H$_2$O) are passively tracked using tracers throughout the mantle evolution, partitioned between the solid and melt phase, and outgassed when melt erupts onto the surface.}
    \label{fig:outgassing_concentration}
\end{figure}
%%% END OF FIGURE 12 %%%
%%%%%%%%%%%%%%%%%%%%%%%
In this work, we track the concentrations of carbon and hydrogen, assuming that they are outgassed as CO$_2$ and H$_2$O. These elements are passively advected with the mantle flow, meaning there are no geochemical interactions between the different species. We further assume a uniform distribution of volatiles throughout the mantle and do not consider preferential trapping of volatiles in specific mineral phases, such as those found in Earth's transition zone, which could limit the transport of water into the lower mantle of a planet \citep{bercovici_wholemantle_2003, Guimond2023}. When melt is produced, volatiles are partitioned between the solid and melt phases according to the partition coefficients described in Section~\ref{section:methods_melting}. 

Figure~\ref{fig:outgassing_concentration} shows the evolution of the mantle's volatile concentration across the different models normalised by the initial concentration $C_0$. After $6.5$\,Gyr, the mantle of model MPlastic has lost $77\%$ of its carbon budget and $72\%$ of its hydrogen budget. However, most of these volatiles ($\approx 50\%$) are lost during the initial $500$\,Myr of the simulation, again reflecting the strong influence of this model's initial conditions. MPlastic includes plastic yielding and the early evolution is dominated by numerous downwellings and extensive melt generation on the nightside, which subsequently cease once the mantle has cooled sufficiently. Models MixCC, MBasal, and MixIntr are easier to compare as they don't include plastic yielding and they all show a similar spin-up phase in the first $500$\,Myr with minimal outgassing, as can be seen from the rather plateau-like curves in the early phases of Figure~\ref{fig:outgassing_concentration}. MBasal, which is only heated from below, outgasses the least volatiles, with around $17$ and $15\%$ of the CO$_2$ and H$_2$O budget being lost. MixCC outgasses around $31$ and $30\%$ of its volatile budget and MixIntr outgasses around $48$ and $46\%$ of CO$_2$ and H$_2$O, respectively. These models are expected to outgas more volatiles compared to MBasal, as they both include internal heating. Here, we assume that internal heating in the mantle is generated from the decay of radioactive sources such as $^{232}$Th, $^{238}$U, or $^{40}$K \citep{Schubert2001}. We further assume that these internal heat sources are constant over time (assuming a present-day Earth-like internal heat production rate) and are not exhausted over time. In reality, because radiogenic heat sources diminish with time, the rates of mantle melting and degassing in the past could have been more than double.

Tidal heating of the mantle could be another important source of internal heating, which we have not considered here \citep{Bolmont2013, Driscoll2015, Hay2019}. Tidal heating could lead to higher rates of volcanism \citep{Peale1979, Seligman2024}. While current observations suggest that K2-141\,b has a negligible eccentricity, and thus minimal present-day tidal dissipation \citep{Barragan2018, Malavolta2018}, tidal dissipation may have been more significant earlier in its history, if its orbit had not yet circularised; or tidal heating may still be episodically enhanced through dynamical interactions with the outer companion, K2-141\,c.  Active volcanism due to strong tidal heating has been proposed as possible source of the observed high mean-molecular weight species (H$_2$S, SO$_2$) in the observed spectrum of the low-density super-Earth L 98-59\,d \citep{Banerjee2024, Gressier2024}, while SO$_2$ has also been suggested in the sub-Earth L 98-59\,b \citep{BelloArufe2025}. In contrast, \citet{Nicholls2025} showed that tidal heating can maintain permanent magma oceans on these planets, from which volatiles are continuously outgassed at equilibrium into an overlying atmosphere. 

Future work could therefore look into how the mantle dynamics and subsequent outgassing is influenced when considering tidal heating or depth-dependent internal heating effects. Additionally, volatile species such as H$_2$O and CO$_2$ can strongly depress the solidus and induce flux melting \citep[e.g.][]{Karato1998, Katz2003, Dasgupta2013}. A future study could investigate how this affects melt production and outgassing rates on tidally-locked lava planets.

From the amount of outgassed volatiles, we can make a rough estimate for the atmospheric pressure build-up, where we neglect any thermodynamic effects (i.e. temperature, condensation), solubilities, or escape. At hydrostatic equilibrium, the atmospheric pressure is given by 
\begin{equation}
P(t) = \frac{M(t)g}{A_{\rm p}}, 
\end{equation}
where $g=21.8$\,m/s$^2$ is the surface gravitational acceleration, $A_{\rm p}$ is the surface area of the planet, and $M$ is the total mass of the outgassed volatile species (CO$_2$ and H$_2$O). 
%%%%%%%%%%%%%%%%
%%% FIGURE 13 %%%
\begin{figure}
    \centering
    \includegraphics[width=0.5\textwidth]{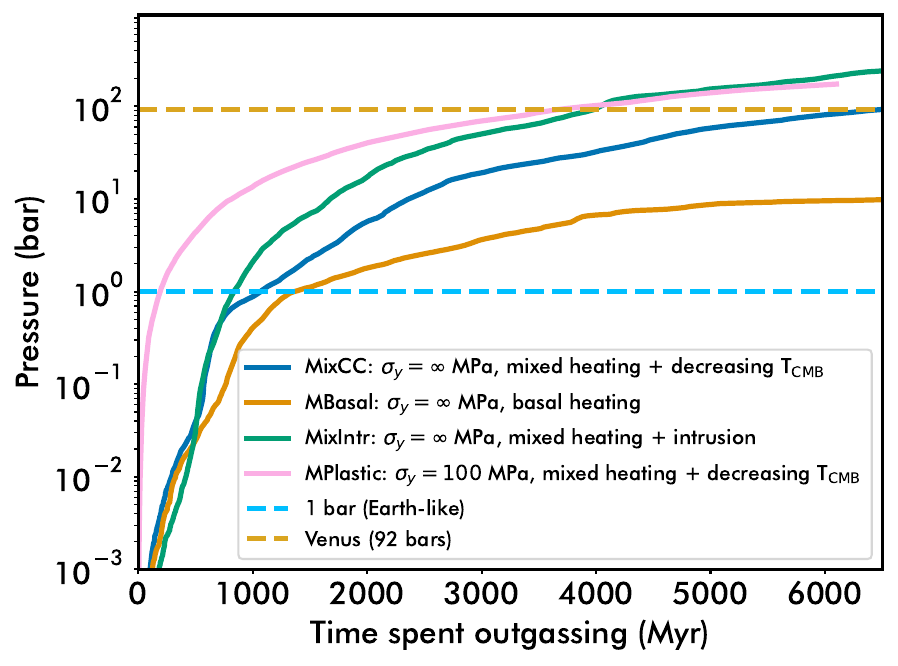}
    \caption{Evolution of total pressure of outgassed volatiles resulting from the cumulative outgassing of CO$_2$ and H$_2$O for the different models. Partial pressures are computed from the mass of each gas species outgassed over time, assuming initial mantle concentrations of $110$\,ppm carbon and $290$\,ppm ppm water. Horizontal dashed lines indicate present-day Earth ($1$\,bar) and Venus ($92$\,bar) atmospheric surface pressures for reference.}
    \label{fig:atmosphere_evolution}
\end{figure}
%%% END OF FIGURE 13 %%%
%%%%%%%%%%%%%%%%%%%%%%%
Figure~\ref{fig:atmosphere_evolution} shows the resulting evolution of the total pressure of the outgassed volatiles. For MPlastic, we again neglected the first $500$\,Myr, during which the model undergoes a spin-up phase characterised by unrealistic high outgassing rates. As expected from the eruption rates (Fig.~\ref{fig:eruption_plot}) and volatile depletion of the mantle (Fig.~\ref{fig:outgassing_concentration}), MBasal exhibits the least volatile outgassing, reaching $109$\,bar after $6.5$\,Gyr. This is a direct consequence of this model only including basal heating, which limits melt production and surface volcanism.

%%%%%%%%%%%%%%%%
%%% FIGURE 14 %%%
\begin{figure}
    \centering
    \includegraphics[width=0.5\textwidth]{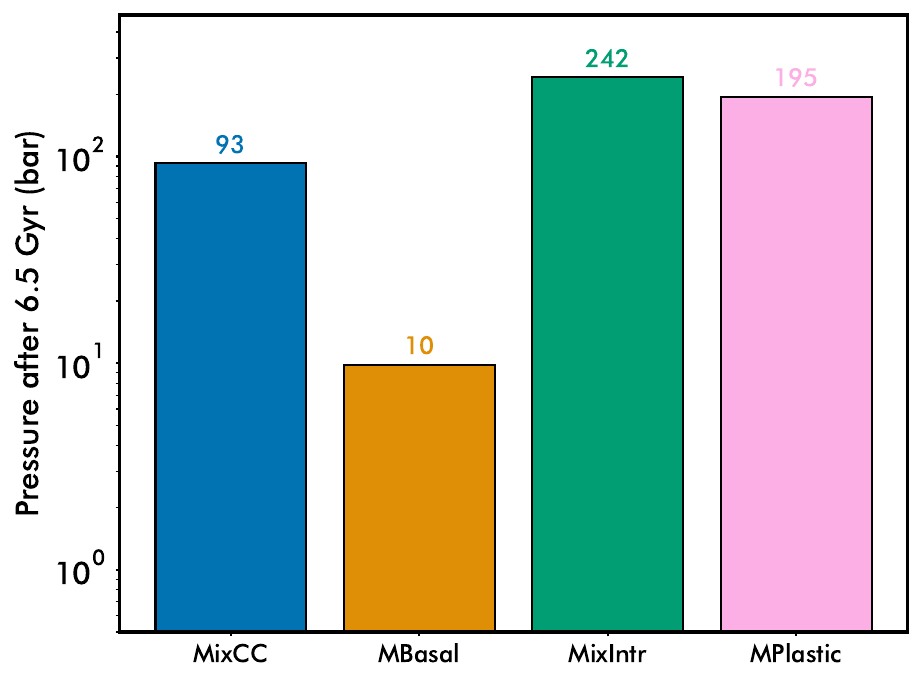}
    \caption{Total atmospheric pressure after $6.5$\,Gyr resulting from the cumulative outgassing of CO$_2$ and H$_2$O for the different models shown in Figure~\ref{fig:atmosphere_evolution}. Each bar represents the total surface pressure calculated from the total outgassed volatile mass assuming initial mantle concentrations of $110$\,ppm carbon and $290$\,ppm water.}
    \label{fig:final_pressure}
\end{figure}
%%% END OF FIGURE 14 %%%
%%%%%%%%%%%%%%%%%%%%%%%

Figure~\ref{fig:final_pressure} shows the total accumulated atmospheric pressure after $6.5$\,Gyr. For MixCC, MixIntr, and MPlastic, the resulting outgassed surface pressures are $93$\,bar, $242$\,bar, and $195$\,bar (the latter extrapolated). MixCC and MixIntr are particularly interesting to compare, as they reach similar mean mantle temperatures after $5$\,Gyr. While MixCC includes a decreasing CMB temperature from core cooling, MixIntr includes intrusion with an eruption efficiency of $30\%$. However, MixCC has a lower eruption rate than MixIntr as can be seen from the total outgassed pressure as well as the total erupted mass (Fig.~\ref{fig:eruption_plot} b). There are two possible mechanisms for this: MixIntr intrudes $70\%$ of the eruptable melt, and one would therefore expect a lower eruption rate for this model. However, since intrusion stores thermal energy in the lithosphere, this can lead to heating and weakening of the surrounding mantle, promoting lithospheric thinning over time. This process facilitates enhanced melt production in later evolutionary stages and resembles the plutonic squishy lid regime, which is expected to arise under low eruption efficiencies and high intrusion efficiencies \citep{Lourenco2018,Lourenco2020}. 
On the other hand, and this might be the dominant mechanism, the decreasing CMB temperature of MixCC reduces the temperature contrast across the lower thermal boundary layer, leading to less vigorous convection and fewer upwellings. This limits the number of upwellings capable of generating melt at the nightside surface, thereby suppressing volcanism. This can also be seen by comparing the snapshots of the mantle temperature at $5$\,Gyr (Fig~\ref{fig:5gyrplot}: MixCC and MixIntr), where MixCC shows fewer and weaker upwellings compared to MixIntr. 

It is important to note that these values represent upper limits on the atmospheric pressure resulting from volcanic outgassing, rather than predictions of actual atmospheric thickness. A significant portion of the outgassed volatiles may be lost to atmospheric escape processes, particularly under the influence of strong stellar irradiation and hydrodynamic escape \citep{Owen2019, Hazra2025, Chatterjee2025}. Given that the $2\sigma$ upper limit of the nightside flux of K2-141\,b is consistent with zero \citep{Zieba2022}, JWST phase curve observations with MIRI/LRS (GO 2347, PI: Dang) and NIRSpec/G395H (GO 2159, PI: Espinoza) are required to determine whether nightside temperatures are cold enough to allow for atmospheric collapse and trapping of outgassed species (e.g., CO$_2$, H$_2$O) on the nightside. Though a thick atmosphere would lead to substantial heat redistribution \citep{Koll-2022-Scaling}, the more likely scenario of a thin rock vapour atmosphere \citep{Nguyen:2024aa} would lead to atmospheric collapse on the nightside in the absence of  stronger nightside surface heat fluxes than that found in our simulations (see e.g., Figure \ref{fig:eruption_conduction_plot}). On the other hand, we assume a zero pressure initial atmosphere, neglecting any primordial envelope or volatiles outgassed from the magma ocean. 
In particular, this calculation only considered outgassing on the nightside. On the dayside, volatiles are expected to be outgassed from (and possibly in-gassed into) the magma ocean \citep{Bower2018, Bower2019, Lichtenberg2021, Maurice2024}, which will further influence their inventory in the planet interior \citep{Dorn2021, Bower2022}.

Outgassing on a tidally locked lava world such as K2-141\,b is therefore likely characterised by two distinct regimes, with magma ocean outgassing on the dayside and volcanic degassing from the nightside, which leads to a pressure gradient and strong winds (on the order of km/s) between the dayside and nightside which could also transport volatiles released on the dayside towards the nightside \citep{Nguyen2020, Nguyen2022}. Importantly, the volatile content of the surface layer of the magma ocean in contact with the atmosphere cannot be sustained indefinitely without replenishment from the magma ocean's deeper layers, as well as from the underlying mantle. \citet{Salvador2023} and \citet{Walbecq2025} have shown that outgassing efficiency from super-Earth magma oceans is limited by their high convective vigour, at least for deep primordial magma oceans. Another recent study has shown that, if a reducing atmosphere is present, oxidation reactions between reducing gases and iron oxides in the magma ocean can generate metallic iron that sinks to the core, leaving behind a buoyant, upper layer \citep{ModirroustaGalian2025}. This compositional stratification would further inhibit volatile transport to the magma ocean surface, where they could outgas. Future work should investigate possible limits to volatile transport within a shallow dayside magma ocean. In our simulations, we find that the dayside magma ocean remains relatively shallow, limiting the total volatile reservoir available at the surface. However, we also observe that plumes on the dayside that rise from the CMB towards the magma ocean are able to deliver fresh material, including volatiles. As a result, outgassing on the dayside is not only dependent on the internal dynamics of the magma ocean, but furthermore by the rate at which fresh volatiles are delivered from the solid mantle. A future study should investigate whether this delivery is sufficient to prevent the magma ocean from becoming desiccated over geological timescales. 

For the initial volatile budget, we assumed a present-day bulk silicate Earth-like volatile budget with 110\,ppm of carbon and $290$\,ppm of water \citep{Hirschmann2018}. However, since modern Earth's mantle volatile content reflects long-term recycling between the surface and interior via plate tectonics, these values may not be representative for all terrestrial planets. Alternative carbon-rich scenarios, such as those proposed for Venus with initial water concentrations of $100-200$\,ppm and initial carbon concentrations of $1000-4000$\,ppm \citep[decreasing to $30-100$\,ppm water and $200-500$\,ppm carbon at present-day;][]{Gillmann2016} could lead to higher pressures of outgassed volatiles due to increased partitioning into the melt \eqref{eq:partitioning}, although we note that none of our simulations result in mantles that have outgassed completely (Fig. \ref{fig:outgassing_concentration}). On the other hand, the volatile budget could also be significantly lower if K2-141\,b underwent strong early atmospheric loss during its primordial magma ocean and high-XUV phases (see Section~\ref{sec:escape}), potentially leading to a volatile-depleted mantle and decreased long-term outgassing if the mantle is simultaneously dessicated. However, the efficiency of this volatile loss would depend on how much volatiles remained dissolved in the magma ocean, as species such as H$_2$O are highly soluble in silicate melts and could therefore be stored in the interior rather than being fully outgassed \citep{Bower2022}.
Our model does not currently incorporate any chemical equilibration or disequilibrium, variations in oxygen fugacity, solubility, cold-trapping of volatiles on the cooler nightside, cloud formation and associated radiative effects, or the thermodynamic effects of surface pressure build-up. All of these factors are expected to influence the speciation, retention, and effective mass of the outgassed atmosphere. For example, if the mantle of K2-141\,b were more reducing than assumed in this study, then its volcanic gas speciation as well as its total outgassing would be affected. Total carbon outgassing would be much less than estimated here because accessory-phase carbon in the melting region would now take the form of graphite, which is not readily melted \citep[e.g.][]{Guimond2021, Brachmann2025}. Any carbon in volcanic gas would take the form of CO and/or CH$_4$ instead of CO$_2$ \citep[e.g.][]{Guimond2021, Brachmann2025}. Consequentially, the atmospheric mean molecular weight would be lower.

Fully capturing the atmospheric evolution of such planets will therefore require models that couple volatile as well as potential rock vapour outgassing  \citep{Zilinskas2023, Piette2023, Seidler2024, Buchem2025} with atmospheric escape \citep{KrissansenTotton2024,Nicholls2025d}, chemistry \citep[e.g.][]{Maurice2024, Brachmann2025}, and supersonic dynamics \citep{Song2025} across the planet's contrasting hemispheres.

\subsection{Atmospheric Escape}
\label{sec:escape}
Ultra-short-period rocky planets are subjected to ionising radiation and stellar winds so intense that volcanic atmospheres may be unsustainable. \cite{Tang2024} demonstrate that a primordial nebular atmosphere on K2-141\,b (5.31 $M_{\earth}$, $2900$\,$S_{\earth}$) would be rapidly lost during post-formation boil-off, unless the planet formed ex situ and migrated inward \citep{2020Millholland}. Furthermore, an exploration of XUV-driven hydrodynamic escape from \cite{Ji2025} found that a $1{,}000$-bar CO$_2$ envelope is stripped away on gigayear timescales in the absence of a magnetic field, placing K2-141\,b towards the airless side of the cosmic shoreline \citep{Zahnle2017} even in highly volatile-rich scenarios. However, the degree to which other factors, namely electron-impact excitation of atomic lines, photochemical controls, and planetary magnetism can protect secondary atmospheres against escape remains poorly understood. 

The gravitational sphere of influence of K2-141 b is greatly reduced by its proximity to the host star --- the L$_1$ Lagrange point lies only about one planetary radius above the surface. Following \citet{Erkaev2007}, the Roche-lobe correction to the gravitational potential is $K = 0.35$, reducing the effective escape velocity by $K^{1/2}$, from 20.5 km s$^{-1}$ to 12.1 km s$^{-1}$. \citet{Nakayama2022} simulated the escape of Earth’s atmosphere under an XUV flux 1000 times the present level and found that the atmosphere remained hydrostatic, reaching a plateau at 1 bar every billion years. In contrast, \citet{Chatterjee2025} showed that transonic hydrodynamic escape at a rate of $6$ bar Myr$^{-1}$ can occur at XUV fluxes of $400\times$ present Earth levels, and suggest that a hydrostatic state becomes unstable when the plasma properties of the upper atmosphere are included. However, the radial ambipolar electrostatic field that promotes escape in ionized conditions is diverted in the presence of closed magnetic field lines via the Lorentz force \citep{Trammell2011}. Thus, particularly for a magnetized planet, estimates of secondary atmosphere lifetimes vary by orders of magnitude, and it remains to be tested whether volcanic outgassing at the rate reported here ($\lesssim 32$ bar Gyr$^{-1}$) could offset escape and sustain an atmosphere.

Is a long-lived geodynamo on K2-141\,b plausible? Multiple modelling studies suggest that the Fe cores of more massive rocky planets are more likely to be liquid, such that thermal convection in this liquid core, as opposed to solidification of a solid inner core, would generate a magnetic field  \citep[e.g.][]{gaidos_thermodynamic_2010, blaske_energetic_2021, white_initial_2025}. The lifetime of this dynamo depends on the cooling timescale of the core \citep[the heat flux from core to mantle staying above some critical threshold;][]{blaske_energetic_2021}. \citet{luo_radiogenic_2024} calculate a magnetic field lifetime easily surpassing 10\,Gyr for a $\sim5\,M_\oplus$ rocky planet with an Earth-like core fraction, unless the mantle viscosity is extremely low and the core is not internally heated, and at 4.5 Gyr this geodynamo could be as strong as 60--80\,$\mu$T.

\section{Conclusions}
In this study, we investigated the mantle dynamics, volcanic eruption rates, and volatile outgassing of the ultra-short period lava world K2-141\,b using the two-dimensional mantle convection code \textsc{StagYY} \citep{Tackley2008}. We investigated how different parameters affect these processes by varying the lithospheric rheology (models with and without plastic yielding), the internal heating mode (basal versus mixed heating), the presence or absence of core cooling effects on the CMB temperature, and the treatment of melt emplacement (with and without intrusion). Our simulations reveal that the hemispheric surface temperature contrast---resulting from tidal locking and inefficient heat redistribution as expected from the Spitzer observations \citep{Zieba2022}---has an important impact on mantle flow, the spatial distribution of melt production and volcanic activity, and the resulting surface composition. 

For the models which do not permit plastic yielding, we find that the mantle stabilises in a degree-2 convection regime, with two downwellings forming near the day-night terminators, at the interface between the magma ocean and the colder, solid nightside. These downwellings enable the recycling of crustal material at the day-night boundary, where it is transported into the mantle. The formation and behaviour of these downwellings is different from classical subduction on Earth, although both result in the recycling of surface material. Hence we add to the growing list of mechanisms theorised to permit lithospheric and volatile recycling on planets without plate tectonics \citep{lenardic_mechanism_1993, zegers_middle_2001, elkins-tanton_volcanism_2007a}. This phenomenon may represent a new tectonic mode characteristic of magma ocean planets with strong day-night temperature contrasts where the nightside remains solid. Notably, this regime combines a strong lithosphere with asymmetric, localised crustal recycling near the boundary with the magma ocean, which emerges without the need for a plastic yielding criterion. 
%%%%%%%%%%%%%%%%
%%% FIGURE 15 %%%
\begin{figure*}
    \centering
    \includegraphics[width=1.0\textwidth]{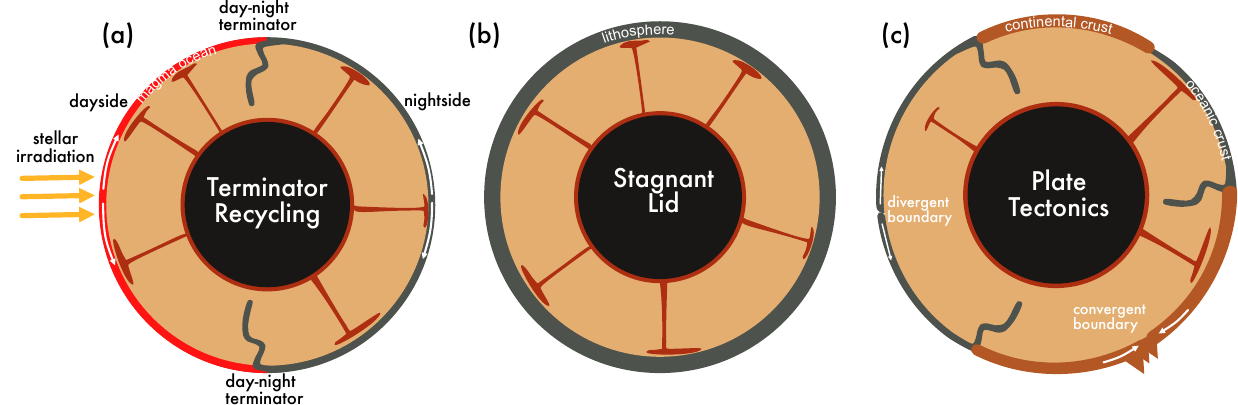}
    \caption{Illustration of different tectonic regimes. (a) Terminator-recycling regime identified in this study, where downwellings form at the day-night terminator at the boundary between the magma ocean and cold, solid nightside. These downwellings enable the recycling of crustal material. (b) Stagnant-lid regime, characterised by a thick, immobile lithosphere. (c) Earth-like plate tectonics, where the lithosphere is divided into several rigid plates that move relative to each other, facilitating subduction zones and crustal recycling.}
    \label{fig:regimes_illustration}
\end{figure*}
%%% END OF FIGURE 15 %%%
%%%%%%%%%%%%%%%%%%%%%%%
Figure~\ref{fig:regimes_illustration}a illustrates the terminator-recycling regime identified in this study, shown in comparison to a stagnant-lid regime, where the thick lithosphere is immobile and prevents efficient recycling of material (panel b), and and Earth-like plate-tectonic regime, characterised by several plates that move relative to each other, with subduction zones at convergent boundaries and seafloor spreading at divergent ones (panel c).

The extent and longitudinal distribution of volcanism on K2-141\,b are strongly influenced by the mode of interior heating and lithospheric rheology. Models with internal heating produce broader and more sustained eruption histories than those heated solely from below. Our models reveal that the nightside heat flux produced by volcanism and mantle cooling is likely challenging to detect using current and near-future space-based telescope observations (Fig.~\ref{fig:eruption_miri_ppm}). However, we find that for some models outgassing is concentrated near the terminator regions (Fig.~\ref{fig:eruption_longitude}), and future studies could explore whether this leads to detectable atmospheric spectral signatures that permit chemical characterisation, and, in the more distant future, even wind speed measurements with sufficient spectral resolution. This could be assessed for JWST in the context of a large program, future missions such as LIFE and HWO, and at high resolution spectroscopy ($R\sim100,000$) for the suite instruments on the upcoming Extremely Large Telescope (ELT) e.g. METIS, ANDES, and PCS. 

We also investigated the build-up of a potential atmosphere over geological timescales and found that continuous volcanism on the nightside can lead to cumulative outgassing of volatiles (e.g. CO$_2$ and H$_2$O), resulting in surface pressures ranging from $10$ to $100$\,bar, assuming an present-day Earth-like volatile inventory (Fig.~\ref{fig:atmosphere_evolution}). For more volatile-rich planets, the resulting outgassed pressures could be substantially higher. This effect could be even more pronounced for younger planets with higher internal heating rates, which could drive more vigorous volcanism. However, these surface pressures may represent upper limits, as our current tracer implementation of tracking volatiles in the mantle is passive and does not account for redox-sensitive processes such as oxygen fugacity-dependent volatile speciation, volatile solubilities, or feedbacks from surface pressure build-up. We also neglect volatile recycling (ingassing), remnant primordial volatile atmospheres, condensation on the nightside, and comprehensive atmospheric thermodynamics. Incorporating these geochemical and thermodynamic processes will be crucial for assessing the stability and composition of a potential atmosphere on K2-141\,b, or other lava worlds. 

In conclusion, our models reinforce the view that lava worlds like K2-141\,b are governed by an intricate interplay between mantle dynamics, magma ocean evolution, and atmospheric processes. Future work should aim to couple 2D (and 3D) interior models with atmospheric chemistry and escape models to assess whether the outgassed volatiles can accumulate and shape detectable atmospheres. 

\section*{Acknowledgements}

TGM and JLB acknowledge the support of the Leverhulme Trust via the Philip Leverhulme Physics Prize. 
This project received support from STFC Consolidated Grant ST/W000903/1 and from the Alfred P. Sloan Foundation under Grant G202114194 to the AETHER project. TGM was supported by the SNSF Postdoc Mobility Grant P500PT\_211044.
RJS has been funded by the Swiss National Science Foundation, Postdoc.Mobility Grant P500PT\_217847. TL acknowledges support from the Netherlands eScience Center (NLESC.OEC.2023.017), the Branco
Weiss Foundation, the Alfred P. Sloan Foundation (AEThER, G202114194), and NASA’s Nexus for Exoplanet System Science research coordination network (Alien Earths, 80NSSC21K0593). CEF acknowledges financial support from the European Research Council (ERC) under the European Union’s Horizon 2020 research and innovation program under grant agreement no. 805445.  CMG is supported by the UK STFC [grant number ST/W000903/1]. RDC thanks UKRI for their support via grant number UKRI1191. A.M. acknowledges support through a studentship provided by the U.K. Science and Technology Facilities Council. Calculations were performed on the AOPP and Astrophysics (Glamdring) HPC clusters. We thank the anonymous reviewer for taking the time to review this manuscript and for their constructive feedback that has greatly improved this work.

%%%%%%%%%%%%%%%%%%%%%%%%%%%%%%%%%%%%%%%%%%%%%%%%%%
\section*{Data Availability}
Selected data for reproducing the figures in this study can be obtained from the Zenodo Data Repository \citep{ZenodoMeier2025}. The mantle convection code \textsc{StagYY} \citep{Tackley2008} is the property of PJT and Eidgenössische Technische Hochschule Zürich (ETHZ) and is available for collaborative studies from Paul J. Tackley. Requests can be made to PJT (\href{mailto:paul.tackley@eaps.ethz.ch}{paul.tackley@eaps.ethz.ch}).
Post-processing of \textsc{StagYY} data was performed using \textsc{StagPy} \citep{Morison2022}, \textsc{NumPy} \citep{Harris2020}, and \textsc{SciPy} \citep{Virtanen2020}. The figures were created using \textsc{Matplotlib} \citep{Hunter2007}. The perpetually uniform colormaps are from \citet{Crameri2018}. 


\bibliographystyle{mnras}
\bibliography{k2-141b_refs} % if your bibtex file is called example.bib

%%%%%%%%%%%%%%%%%%%%%%%%%%%%%%%%%%%%%%%%%%%%%%%%%%

%%%%%%%%%%%%%%%%% APPENDICES %%%%%%%%%%%%%%%%%%%%%

%\appendix


%%%%%%%%%%%%%%%%%%%%%%%%%%%%%%%%%%%%%%%%%%%%%%%%%%

% Don't change these lines
\bsp	% typesetting comment
\label{lastpage}
\end{document}